\documentclass[aps,prl,preprint,groupedaddress]{revtex4-2}

\usepackage{amsmath}
\usepackage{mathtools}
\usepackage{graphicx}

\begin{document}


\title{Black hole information turbulence and the Hubble tension}


\author{
	Juan Luis Cabrera Fernández
}
\affiliation{
Departamento de F\'isica Aplicada, ETSIAE, Universidad Polit\'ecnica de Madrid, Pza. Cardenal Cisneros 3, Madrid \& 28040, Spain	\\
and\\
Laboratorio de Din\'amica Estoc\'astica, Centro de F\'isica, Instituto Venezolano de Investigaciones Cient\'ificas, Caracas \& 1020-A,  Venezuela.
}


\date{\today}

\begin{abstract}
A major  outstanding challenge in cosmology is
the persistent discrepancy between the Hubble constant 
obtained from early and late universe measurements -- the Hubble tension.
Examining cosmological evolution through the lens of information growth within a black hole 
we show the appearence of two fractal growing processes 
characterizing the early 
and late ages. 
These fractals induce space growth rates of
   $(62.79\pm5.59)$  km/s/Mpc and   $(70.07\pm0.09)$  km/s/Mpc; 
 close to the current values of the  Hubble constants
 involved in the tension. 
 These results strongly suggest that the Hubble tension is not given by 
  unexpected large-scale structures or multiple, unrelated errors 
but  by innate properties 
underlying the universe dynamics.
\end{abstract}


\maketitle

Main astrophysical experiments have improved the accuracy of 
measurements of the universe's expansion rate -- the Hubble constant,   $H_0$ --, 
 derived from early universe measurements 
 from Cosmic Microwave Background experiments, 
or from late time based on local measurements of distances and redshifts, 
e.g.,  from experiments with pulsating Cepheid variables or type Ia supernovae. 
The  persistent discrepancy between the Hubble constant value 
obtained from these two different approaches -- the 
so called Hubble tension \cite{DiValentino} -- is an outstanding challenge.
The remarkably precision and consistency of the data 
impose stringent constraints on potential solutions, 
calling for a hypothesis robust enough to account for diverse observations, 
that may even involve  novel physical phenomena.

Here we analyzed this problem from the point of view of the physics inside a black hole. 
Let's begin considering that the inside of a black hole has been described  as a quantum circuit where the 
evolution 
of   $s$ qubits,  in a space of states of   $K$ qubits, obeys   \cite{s1}, 
\begin{equation}
\Delta s = s_{\tau+1} - s_{\tau}\ = \dfrac{(K-s)s}{K-1}, \label{e1}
\end{equation}
\noindent
where an average number of new infections,   $\Delta s$,  
are produced in the circuit next step   $\tau + 1$ 
\footnote{Using the notation   $s(\tau+1) = s_{\tau+1}$ and   $s(\tau)=s_\tau$, with   $\tau$ integer.}.
\noindent
Previous studies  \cite{s1} turned the iteration time,  
$\tau$, into a continuous variable and replaced  
the difference equation by a differential equation.
However, if one sticks to the original discrete time step, it is easily shown that Equation (\ref{e1})  
is equivalent to the logistic map \cite{k1} (see    the Appendix ), 
\begin{equation}
s_{\tau+1} =  r s_\tau ( 1- s_{\tau}) \label{logmap},
\end{equation}
with a control parameter, 
$r \equiv   \dfrac{2K-1}{K-1}   $.
Thus we can write,   $K(r)= \dfrac{r-1}{r-2}$, 
meaning that there are a large number of qubits when   $r \sim 2$.
One may expect something relevant to happen  
with a large population of qubits that corresponds with a large entropy situation \cite{s1}, 
but the logistic map with   $r \sim 2$, 
simply converges to the fixed point   $\sim \left(1-\dfrac{1}{r} \right)$. 

\section{Black hole information dynamics with limited resources}
\begin{figure}[htb]
\includegraphics[width=0.6\textwidth]{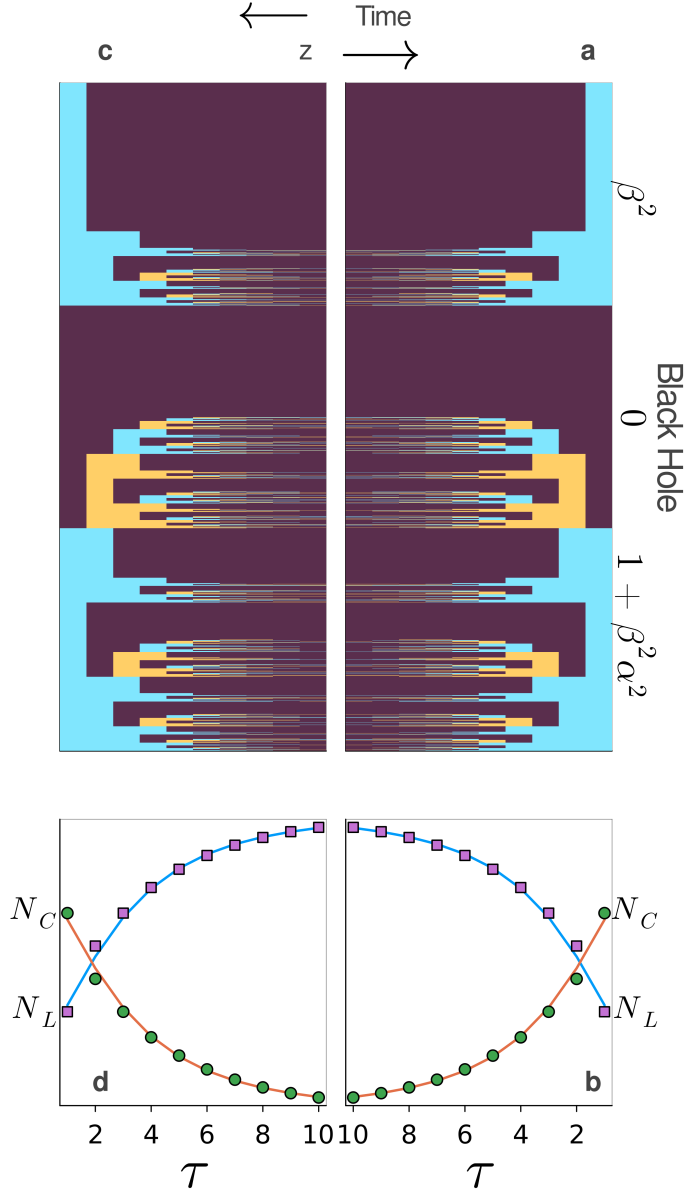}
	\caption{ 
			{\bf Fractal turbulent cosmology.} 
			({\bf A}) The coefficients describing the evolution of a population of qubits inside the black hole form a fractal cascade as the iterations   $\tau$ increase 
			(described from right to left).  
			({\bf B})  The fraction of lacunar zero coefficients (squares)   $N_L$ increases 
			while the fraction of coloured non-zero coefficients (circles)   $N_C$
			decreases describing a cantor-like dust
			in ({\bf A}), 
			while   $n_c$ grows exponentially (see    the Appendix ).
			({\bf A})  was calculated with   $r=r_H$, after iterating 
			$\tau_f=10$ generations, giving rise to 
			$3^{10}=59049$ coefficients. 
			Colours describe the coefficient's intensities in the nested cascade 
			(See \cite{jlc} and    the Appendix )
			Non-zero coefficients are coloured blue and brown and the lacunar component is violet. 
			({\bf C})  Once the cascade stops, it is inverted and 
			({\bf D})    $N_C$ increases while   $N_L$ decreases reaching the initial proportions 
			$\dfrac{2}{3} \sim 0.666$ and   $\dfrac{1}{3} \sim 0.333$. 
			In ({\bf B})  and ({\bf D}) lines are a best fit of the   $\Lambda$CDM model 
			with parameters    $C_z = 0.37626$,   $p_k = 0.000216$,   $p_m = 0.167422$ and 
			$p_r =  -0.000219$ (see    the Appendix ) 
			and \cite{jlc} for the meaning of   $\alpha$ and   $\beta$.
\label{fig1}}
\end{figure}

Nevertheless, one should note that the number of qubits  in the  black hole   $K$-space  is not infinite. 
So,  the number of qubits states available in a given generation
must be regulated 
by the number of states left available by previous  generations. 
This is an argument frequently used in simple population dynamics in the presence of limited resources.
A straightforward approach consist of modelling the progressive reduction of resources with a regulation term determined by the preceding generation. 
Thus the next generation is both, 
proportional to the current population,   $s_{\tau+1} \sim s_\tau$, 
and to the regulation term   $s_{\tau+1} \sim ( 1- s_{\tau-1})   $. 
With these considerations in mind Equation (\ref{e1}) takes the form of 
the Delayed Regulation Model (DRM)  \cite{maynardSmith68}.
\begin{equation}
s_{\tau+1} =  r s_\tau ( 1- s_{\tau-1}) \label{e2}.
\end{equation}
This map has a Hopf bifurcation at    $r_H \equiv2$ and as Eq. (\ref{logmap}) also leads to chaos 
\cite{PounderRogers80,aronson82,morimoto88}. 
Notably, in the frequency domain the DRM can be deduced from the incompressible 
form of the Navier-Stokes equation. 
At any generation   $\tau$, the leading terms of the eigenvalues governing the dynamics are second-order polynomials whose coefficients are themselves nested second-order polynomials, with this nesting repeated   $\tau$ times. The recursive generation of these coefficients can be described precisely \cite{jlc}, enabling the construction of an analytical cascade that exhibits the   $-\frac{5}{3}$ scaling law characteristic of the power spectrum in isotropic homogeneous turbulence. 
The cascade's first steps  are depicted in the  Fig.   \ref{Sfig4}. 
This cascade can be represented as the fractal 
seen in Fig. \ref{fig1}A, 
where the value of the three coefficients  for the 
initial condition 
(second order:    $1+\beta^2 \alpha^2$, linear:    $0$ and independent:    $\beta^2$, terms)
are mapped to  segments of size    $\dfrac{1}{3}$ on the unit interval, 
and  subsequently subdivided 
following the coefficients generating rule 
(see details in     the Appendix and     \cite{jlc})
From now on, the generated fractal will be considered a sufficiently valid course grain approximation  
of the turbulent cascade in the frequency domain.
As for   $K$ large enough  
the control parameter is close to the Hopf bifurcation value,    $r \sim r_H$, 
the population of qubits inside the black hole describes a critical dynamics. 
After a few steps   $\tau$,  
the coefficients governing  
the qubits population reach a 
fully developed turbulent state, 
i.e., a state of information turbulence in a space partitioned in 
-- the fast growing number --    $3^{\tau}$ states.  
 After a probably large, but finite number of iterations, 
 the population of qubits should exhaust its resources 
 in the black hole   $K$ bounded space. 
  As a result, after such a large    $\tau$ 
  -- much larger than the exemplified in Fig.    \ref{fig1}A -- 
  no additional incoming energy will feed the cascade and one may expect turbulence to recede from its fractal limit set. 
  A distant observer of such a receding situation would 
  see the same cascade backwards. 
 However, such an observer would be unable to witness 
 the final fully developed cascade, but would rather see 
 a later stage with turbulence already receding, i.e., 
 this later state acts as an horizon. 
 Fig.  \ref{fig1} illustrates this process:
 the direct cascade happening inside the black hole (Fig.  \ref{fig1}A)
 is mirrored by the inverse cascade (Fig.  \ref{fig1}C).
 
 \section{Measuring the fractal cascade}
 To  quantify the evolution of the cascade components during both, 
 the direct and the recession stages, at a given generation,    $\tau$, 
 we count the number of coloured,    $n_C(\tau)$, and lacunar,    $n_L(\tau)$, 
components in the fractal. 
With these quantities we can calculate the 
normalised number of coloured non-zero coefficients,    $N_C(\tau)$, and 
of lacunar zero coefficients,    $N_L(\tau)$. 
With increasing iterations,    $n_C(\tau)$ and    $n_L(\tau)$ grow exponentially 
(see Fig. \ref{Sfig5}), while    $N_C(\tau)$ and     $N_L(\tau)$ 
describe the curves seen at Figures \ref{fig1}B and \ref{fig1}D. 
During the recession, in the long run,    $N_C$ dominates over    $N_L$. 
It is striking the similarity of the inverse cascade's 
   $N_C(\tau)$ and     $N_L(\tau)$ 
with the evolution of the fractional energy density 
of nonrelativistic matter 
and dark energy components 
\footnote{Note that in this construction the total energy density is always conserved and equals    $N_C + N_L =1$.}.
To better quantify such a similarity we  
fitted seven different cosmological models 
\cite{doran2001,doran2006,poulin,shi2012} in terms of the redshift 
   $z$ (See Appendix)
and found that the 
   $\Lambda$CDM  and two Early Dark Energy models 
produced  better results. 
We included the parameter   $C_z$ to establish a relationship 
 between   $\tau$  \footnote{Rindler time according to \cite{s1}.}
 and   $z$, i.e., 
    $z = C_z \tau$.
These mentioned three models yielded very similar   $C_z$'s (Table  \ref{table1}).
Hence, very large redshifts correspond to the  
origin of the cascade and "just large" ones to a time dominated by turbulent states with large   $\tau$.
According to this description the backward turbulence progression 
is able to roughly describe 
the universe evolution from the instant when the original turbulence 
- developed by a population of qubits inside a black hole -, started to recede. 
It must be stressed that in this framework we are dealing with information turbulence, 
whose relation with a barotropic cosmological perfect fluid is, to the best of our knowledge, currently unknown. 

\section{How the fractal space growths}
There is further evidence supporting the present description.
The cascade is formed by non-zero coefficients (called coloured) and zero ones (called lacunar). While growing, the fractal structure generates an space where both components are intertwined. 
There is no structure if both components are not present.
A measure of the space filling characteristic of the fractal is its dimension. 
In particular, we may consider two different assertions of this quantity: 
a course grain dimension,   $D_{cg}(\tau)$
\footnote{Course grain in the sense specified above.}, 
and a motif dimension,   $D_m(\tau)$.

Calling the first two non zero coefficients,   $a$ and   $b$, 
and the four non zero coefficients in the next generation,  
$a$,   $b$,   $c$ and   $d$ (see Fig.  \ref{Sfig4}) the full fractal structure 
-- without using information about the coefficients intensities given by the generating rule Eq. (\ref{recurrence})--  
can be recreated following rules:
\begin{align}
	ab \rightarrow\,\, &abcd , \label{r1}\\
	abc \rightarrow\,\,  &abcd , abc, \label{r2}\\
	abcd \rightarrow\,\, &abcd , abc, ab. \label{r3}
\end{align}   
Any motif in the left side generates in the next generation the motifs at the right side. 
These rules greatly simplify the calculations and is particularly useful when trying to establish the fractal dimension as it allows us to reach a higher   $\tau_f$. Let's define an approximation to the course grained dimension as, 
\begin{equation}
D_{cg}(\tau) \equiv \dfrac{\log  n_{C}(\tau) }{ \log 3^{\tau}}.
\end{equation} 
One expects that for   $\tau$ large enough this quantity must converge to the fractal dimension of the limit set, i.e.,  
$D_{cg} \equiv \lim_{\tau \rightarrow \infty} D_{cg}(\tau)$. 
As we are unable to perform a calculation for a very high   $\tau_f$, 
we settle for iterating until    $\tau_f = 27$, as shown in Fig.  \ref{Sfig6}A. 
It can be seen that   $D_{cg}(27) \sim 0.7288$, 
and that the increments defined by,
\begin{equation} 
\Delta D_{cg}(\tau) \equiv D_{cg}(\tau+1) - D_{cg}(\tau),
\end{equation}
diminish considerable with   $\tau$, as   $\Delta D(26) \sim 0.0003$ 
(Fig.  \ref{Sfig6}B).
Therefore, for    $\tau$ large enough,   $D_{cg}(\tau)$ is 
a measure of how the coloured component of the fractal fills an embedding  space that growths as   $3^\tau$. 
Let's calculate the ratio,   $\dfrac{D}{\Delta D}(\tau)$,  and 
remark that for the inverse cascade shown in the main text Fig. \ref{fig1}C, 
it is satisfied that 
$D_{cg\,{I}} = D_{cg}^{-1}$ and   $\Delta D_{cg\,I} = \Delta D_{cg}^{-1}$ 
($I$ denoting the inverse cascade). 
Consequently,  
 \begin{equation}
 	\left(\dfrac{\Delta D_{cg}}{D_{cg}}\right)_{I} = \dfrac{D_{cg}}{\Delta D_{cg}}, \label{dcg}
 \end{equation}
 This function is shown at Fig.  \ref{Sfig6}C which is qualitatively 
 similar to the evolution of the Hubble parameter   $E(z)$ 
 \cite{Salahedin}.
 Eq. (\ref{dcg}) describes how the coloured component of the cascade growths 
 while taking into account the grained structure of the geometrical object. 
 One may also consider an additional measure even courser that   $D_{cg}$ by 
 measuring the number of motifs as defined at the left side of the rules given by 
 the equations (\ref{r1})-(\ref{r3}). Let's define an approximation to a motif dimension as 
 \begin{equation}
 	D_{m}(\tau) \equiv \dfrac{\log  n_{motif}(\tau) }{ \log 3^{\tau}},
 \end{equation}
where,   $n_{motif}(\tau)$, is the number of motifs forming the fractal set at iteration   $\tau$, 
without distinguishing between different motifs. 
We expect that this quantity shall converge to the motif dimension, i.e., that   $D_{m} \equiv \lim_{\tau \rightarrow \infty} D_{m}(\tau)$. 
Furthermore, we can also define the respective successive differences by, 
\begin{equation} 
	\Delta D_{m}(\tau) \equiv D_{m}(\tau+1) - D_{m}(\tau),
\end{equation}
and proceed to calculate how -- in the inverse cascade -- the structure at the level of the motifs grows,  
 \begin{equation}
	\left(\dfrac{\Delta D_{m}}{D_{m}}\right)_{I} = \dfrac{D_{m}}{\Delta D_{m}}. \label{dm}
\end{equation}
The behaviour of   $D_{m}(\tau)$,   $\Delta D_{m}(\tau)$ and   $\left(\dfrac{\Delta D_{m}}{D_{m}}\right)_{I}   $ is illustrated in 
the Fig.  \ref{Sfig7}C 
and, as for the case of   $D_{cg}$, it also shows qualitative similarity with   $E(z)$. 
In the following   $x$ will be used to denote   $cg$ or   $m$. 
Let's remark that  for the inverse cascade shown Fig.  \ref{fig1}C  it is satisfied that  $D_{{x}_{I}} = D_x^{-1}$ 


\section{Determining the Hubble constants} \label{SM7} \hfill\\
To analyse the observed similarity between the behaviour of eqs. (\ref{dcg}) and (\ref{dm}), and   $E(\tau)$, we found it convenient to fit   $y(\tau) \equiv \left(\dfrac{\Delta D_{x}}{D_{x}}(\tau)\right)^2_{I}$ to, 
\begin{equation}
	 \mathcal{M}(\tau) \equiv (p_h E(\tau))^{p_e}.   \label{hm1}
\end{equation} 
Here,   $p_h$ and   $p_e$ are fitting parameters, measuring proportionality and the scaling exponent,  respectively. 
The analysis was restricted to the best behaviour models as shown above. 
It is possible to find a set of parameters   $(p_h,p_e)$ yielding satisfactory 
fits  such that 
$y(\tau) \sim \mathcal{M}(\tau)$. 
The  fitted curves are depicted in the left columns 
of the Figure \ref{fig3} and  Figures \ref{Sfig8} and \ref{Sfig9}
the corresponding parameters values are summarised in the  Table  \ref{table2}.
From these results it is satisfied that, 
\begin{eqnarray} 
	(p_h E(\tau))^{p_e} \sim \left(\dfrac{\Delta D_x}{D_x}(\tau)\right)^2_{I},  \nonumber \\  
	p_h^{p_e} \left(\dfrac{H(\tau)}{H_0}\right)^{2p_e} \sim \left(\dfrac{\Delta D_x}{D_x}(\tau)\right)^2_{I},
\end{eqnarray}
and we can define,
\begin{equation} 
	\Upsilon(\tau)  \equiv  \dfrac{p_{h}^{1/2}}{ \left(\dfrac{\Delta D_x}{D_x}(\tau)\right)_{I}^{1/ p_e}  } H(\tau). \label{U11}
\end{equation}
We would like to substitute   $H(\tau)$ by the posterior fitting. 
To do so we assume  that the relative rate of change of the cosmic scale factor,   $a$, can be approached by the relative rate of change of the fractal dimension, such that,
\begin{equation} 
	H(\tau) = \dfrac{\dot{a}}{a} \sim 
\left(\dfrac{\Delta D_x}{D_x}(\tau)\right)_{I} 
\sim \left[ (p_h E(\tau))^{p_e} \right]^{1/2} = \mathcal{M}(\tau)^{1/2}  \label{h1}
\end{equation}
Let's note that raising to one-half is equivalent to 
rescale $p_e \rightarrow p_{e}^{\prime} \equiv \dfrac{p_e}{2}$, then, 
\begin{align} 
 (p_h E(\tau))^{p_e} \rightarrow   (p_h E(\tau))^{p_{e}^{\prime}},\,\,\, \text{and}\\
	\left(\dfrac{\Delta D_x}{D_x}(\tau)\right)^{1/ p_e  } _{I} 
\rightarrow 
\left(\dfrac{\Delta D_x}{D_x}(\tau)\right)^{1/2{p_{e}^{\prime}} }_{I}.
\end{align}
Substituting (\ref{h1}) into (\ref{U11}) we obtain,
\begin{align} 
\Upsilon(\tau) &\rightarrow
\dfrac{p_{h}^{1/2}}{ \left(\dfrac{\Delta D_x}{D_x}(\tau)\right)_{I}^{1/{2p_{e}^{\prime}}}  } ( p_h E(\tau))^{p_{e}^{\prime}}\\
	 	 &= \dfrac{p_{h}^{1/2}}{ \left(\dfrac{\Delta D_x}{D_x}(\tau)\right)_{I}^{1/2p_{e}}  }  \mathcal{M}_1(\tau)^{1/2}, \label{U1}
\end{align}
where we have omitted the prime in the last expression. 
To calculate the Hubble constant we make the ansatz, 
\begin{equation} 
	H_0  \sim  \lim_{\tau \rightarrow 1} \Upsilon(\tau), \label{limit}
\end{equation}
were   $p_e$ is the value got from the fitting --  doubled to keep its original effect -- and   $\tau \rightarrow 1$, as it is our current time 
in the fractal formulation. 
The results calculated for   $H_0$ are summarised in Table   \ref{table2}, and the representation of the equation
(\ref{U1}) is shown in the right columns of the Figure \ref{fig3} and 
 Figures  \ref{Sfig8}  and \ref{Sfig9}. 
$H_0$ was extrapolated to   $\tau=1$ using 
the Julia package   \textsc{Interpolations.jl}.
To calculate an error estimation for   $H_0$ we took the mean value of the 
propagated error of the equation 
(\ref{U1}) (see the Appendix final section).

We found that   $\Lambda$CDM was the only model producing two results compatible with current accepted values. 
These two Hubble constants are, 
\begin{equation}
	H_{0_{cg}} \sim (62.79\pm5.59) \,\, \text{Km/s/Mpc},
\end{equation}
\begin{equation}
	H_{0_m} \sim (70.07 \pm0.09)\,\,  \text{Km/s/Mpc},
\end{equation}
as    $\Upsilon(\tau)$
is in units of   $H(\tau)$. 
It is important to highlight that   $H_{0_{{cg}}}$
is associated with the filling of the space of the dust grains forming the fractal limit set, as measured by   $D{cg}$, while   $H_{0_{m}}$ is associated with the filling of the space of the larger structures formed by the fractal motifs, as measured by   $D_m$.
These different ways of determining the Hubble constant open the door to an 
innovative and plausible explanation of the fractal origins of the Hubble tension, 
as measures based on the Cosmic Microwave Background (CMB) 
experiments are associated with a description of the early Universe at   $z > 1000$ 
\cite{DiValentino} where, the grained detail of the fractal determines the result. 
Meanwhile, measures based on "shorter" local distances are not influenced by such a grained detail but by the even courser profile of the 
structure given by the fractal motifs. 
Remarkably,   $H_{0_{cg}}$ is compatible with 
mostly all the determinations of   $H_0$ based in CMB, e.g.,  
according to \cite{DiValentino}, 
the most widely cited prediction from Planck in a flat 
$\Lambda$CDM model is 
$H_0 = (67.27 \pm 0.60)$ km/s/Mpc \cite{Aghanim}, which is in the range of   $H_{0_{{cg}}}$. 
Meanwhile, the obtained value for    $H_{0_{m}}$ 
approximates certain estimations of the Hubble constant based on local universe measures, 
e.g.,   $69.8$ \cite{jang2017},   $ 69.6$ \cite{freedman2020},   $70$ \cite{ hoyt2021} and   $70.5$ \cite{khetan2021},
where for simplicity we have intentionally omitted the reported errors.

An interesting result from calculating   $H_0$ with the EDE and EDEP models is that 
they are able to produce "acceptable" values just for the 
$H_{0_{{cg}}}$ case:   $(78.91\pm 0.91)$ km/s/Mpc and   $(73.55\pm 2.75)$  Km/s/Mpc, respectively; 
but produce undervalued results for   $H_{0_{D_m}}$:  
$(37.57\pm 0.69)$ km/s/Mpc and   $(36.91\pm 0.53)$ Km/s/Mpc,  respectively 
(Table   \ref{table2}).
This situation may be indicating that these models are limited in their 
description at shorter redshifts, but are approximately 
capturing features of the early universe, particularly in the EDEP model case.

\section{Final considerations}
We have seen that the population of qubits evolves in a critical state 
where the system shows phases of all sizes.  
In this context, one may interpret phases as subsets of connected gates and 
advance the conjecture that, in such a critical state, the quantum circuit 
underlying the full dynamics described in the Fig.  \ref{Sfig4},  
would be able to execute any computational task.  
In other words, for   $K \rightarrow \infty$, the resulting circuit would have 
an unlimited computational power. 
Inside the event horizon, a quantum circuit as proposed by \cite{s1}
would have such a characteristic. 
Meanwhile, the leading term in the eigenvalues that describes the population of qubits 
works as an envelope containing a set of differentiated coefficients 
that mimics the evolution of both, the dark energy and the matter densities in the universe. 
It must be stressed that the coefficients' magnitude \cite{jlc}
have been used to generate the fractal, but has not played any additional role,  
as the subsequent analysis has been restricted to counting the cascade's components 
as given by Eqs. \ref{r1}-\ref{r3}.
The consideration of magnitudes may also unveil unexpected outcomes, and  
 aspect left for future works.  

This research shows how inside black hole physics, information and turbulence 
seem to be intertwined with each other in a cosmological framework offering a plausible explanation to the Hubble tension based on 
"simple" systems nonlinear dynamics \cite{k1}.
However it offers no clues about  how a pure qubits evolution results in observables that coincide in an acceptable degree with capital astrophysical determinations. 
We have left for a future work the treatment of the $S8$ tension - $S8$ measuring of how inhomogenous is the universe -, which 
at first sight seems to be also explainable in the current two fractal framework.

\begin{acknowledgments}
The author gratefully thanks Nuria Alvarez Crespo (U-TAD) for comments and discussions 
and  Daniel Duque Campayo (UPM) and María Angeles Moliné (UPM) for their comments at the initial stage of this work; 
also  thanks  the community of developers of Julia Lang.
The author acknowledges personal support from Prof. M. C. Pereyra (UNM) and 
 the use of Google, Perplexity.ai and Claude.ai for fast searches of sources of information.  
 This research received no financial support from any agency. 
\end{acknowledgments}


\appendix
\noindent
\section*{Appendixes}

\subsection*{Derivation of the logistic equation from Eq. (\ref{logmap}) } \label{SM1} \hfill\\
From the equation (\ref{e1}) 
with   $\Delta s = s_{n+1} - s_{n}$, one obtains,
\begin{eqnarray}  
s_{n+1}  s_n +   \dfrac{K  }{K-1}  s_n -  \dfrac{s_n^2  }{K-1}   \label{e4}, 
\end{eqnarray} 
\noindent
or,
\begin{eqnarray}  
s_{n+1}   \left( \dfrac{2K -1 }{K-1} \right) s_n \left( 1 -  \dfrac{s_n  }{2K-1}  \right) \label{e5}.
\end{eqnarray} 
\noindent
Applying the transformations:
\begin{eqnarray}
\dfrac{s_n}{2K-1} &\rightarrow& s_n^{\prime} \nonumber\\
s_n &\rightarrow& s_n^{\prime } (2K-1) \nonumber\\
\dfrac{s_n}{K-1} &\rightarrow& s_n^{\prime}  \left(  \dfrac{2K-1}{K-1} \right)   \nonumber,
\end{eqnarray}
 equation (\ref{e5}) is,
\begin{eqnarray}
s^{\prime}_{n+1} (2K-1)  =    \dfrac{(2K-1)^2}{K-1}   s_n^{\prime} ( 1- s^{\prime}_{n}) \label{e6}, 
\end{eqnarray}
that after omitting primes and defining the control parameter, 
$r \equiv \left(  \dfrac{2K-1}{K-1} \right)$, 
allows us to obtain  the well known logistic map: 
\begin{eqnarray}
s_{n+1} =  r s_n ( 1- s_{n}) \label{e7}.
\end{eqnarray} 
A brief but excellent account on the complex behaviour of Eq.  (\ref{e7}) was written by 
Leo Kadanoff in \cite{k1}. The logistic map undergoes a sequence of bifurcations   characterized by the universal Feigenbaum constant \cite{feigenbaum}, leading to chaos.

\section{An illustrative example of the leading term at generation   $\tau = 3$ of the eigenvalues governing the dynamics of the population of qubits} \label{SM2}

The leading term at generation   $\tau = 3$ is called here   $\Lambda_3$. 
We can grasp how the eigenvalues behave by observing the way the
term   $\Lambda_3$ is assembled (see \cite{jlc} for the meaning of   $\alpha$ and   $\beta$),
\begin{equation} 
\Lambda_3  \equiv \left\{ 
\begin{array}{c}
\left[ 
\begin{array}{c}
\left( 1+\beta ^2\alpha ^2\right) { r_1}^2
+\left(\beta ^2+\beta ^4\alpha ^2\right) 
\end{array}
\right] { r_2}^2\\
+\left[ 
\begin{array}{c}
\left(
-2\,\alpha \,\beta ^2-2\,\beta ^4\alpha ^3\right) { r_1}^2
\end{array}
\right] { r_2}\\
+\left[ 
\begin{array}{c}
\left( \beta ^4\alpha ^2+\beta ^6\alpha ^4\right) { r_1}^2
\end{array}
\right] 
\end{array}
  \right\} { r_3}^2 \nonumber
  \end{equation}
\begin{equation} 
+\left\{ 
\begin{array}{c}
\left[ 
\begin{array}{c}
\left( -2\alpha \,\beta ^2\right) \,{ r_1}^2\\
+\left(-2\,\alpha \,\beta ^4\right) 
\end{array}
\right] { r_2}^2\\
+\left[ 
\begin{array}{c}
\left( 2\,\alpha^2\,\beta ^4\right) { r_1}^2
\end{array}
\right] { r_2}
\end{array}
  \right\} { r_3}  \nonumber
  \end{equation}
\begin{equation} \label{A3}
+\left\{ 
\begin{array}{c}
\left[ 
\begin{array}{c}
\left( \beta ^4\alpha ^2\right) { r_1}^2\\
+\left( \alpha^2\beta ^6\right) 
\end{array}
\right] { r_2}^2 
\end{array} 
\right\}
\end{equation}
The nested coefficients structures a cascade as illustrated in Fig.  \ref{Sfig4}.
Given the initial coefficient values for 
the second order term:   $C_2= 1+\beta^2 \alpha^2$, the linear term :   $C_1=0$,  
and the independent term:   $C_0=\beta^2$; the generation rule
allow us to calculate the coefficients that will form the leading term of the eigenvalues at the next iteration   $\tau=2$, 
shown in the second row of Fig.  \ref{Sfig4}.
Now, this new set of coefficients allows for the calculation of the nested coefficients
of the leading term of the eigenvalues at the iteration   $\tau=3$ (third row at  the same figure); 
a full cascade is generated 
by iteratively applying the generation rule. 
On each iteration   $\tau$,   $3^{\tau}$ new terms are generated. 

\subsection*{The generation rule} \label{SM3}

The coefficients of higher order values of the leading term can be precisely obtained from 
the preceding terms by the following generation rule. 
In general, if we know   $\Lambda_1$ and   $\Lambda_2$,
we can obtain   $\Lambda_{\tau}$ given that, in the generation   ${\tau}-1$, the 
term   $\left\{ C_{2} r_{i}^2 + C_{1} r_{i} + C_{0}   \right\}
r_{i+1}^{k}   $, 
with   $k=0,1,2$;  generates the polynomial
\begin{equation}
\left\{  
\begin{array}{c}
\left[ C_{2} r_{i}^2 + C_{1} \chi_1 r_{i} + \beta^2 C_{2} \right] r_{i+1}^{2} \\
+ \left[  -k \alpha \beta^2 C_2 r_i^2 + C_1 \chi_2  r_i + C_1 \chi_3   \right] r_{i+1} \\
+ \left[ \beta^2 \alpha^2 C_0 r_i^2 + C_1 \chi_4 r_i + C_1 \chi_5 \right]  
\end{array}
\right\} r_{i+2}^{k} \label{recurrence},
\end{equation}
\noindent in the next generation   ${\tau}$. There is no need to determine the
unknowns,   $\chi_1, \chi_2, ... , \chi_5$,
because in the current situation    $C_1 = 0$
 (see details in \cite{jlc}). 

\subsection{Fractal construction} \label{SM4}

The structure of the   $\Lambda_{\tau}$'s  is better represented by the induced fractal it generates: 
to the initial coefficient values   $C_2$,   $C_1$ and   $C_0$, 
same size segments on the unit interval are assigned. 
After repeated iterations  
each subset is divided by a factor of   $3$, and
the newly generated coefficients obtained with the generation rule updates
the new   $N_{\tau}=3^{\tau}$ subintervals.
A detailed account is available at \cite{jlc}.

\subsection{Fitting the cascade's components} \label{SM5}

Fig.  \ref{Sfig4} shows the number of coloured   $n_C$ and lacunar   $n_L$ components in the 
inverted cascade. Both quantities grow exponentially following fits:
\begin{equation}
n_C(\tau)   \sim  0.812 \,e^{0.807\,\tau}
\end{equation}
\begin{equation}
n_L(\tau)  \sim  0.561 \,e^{1.142\,\tau}
\end{equation}
One may try to fit the fraction of the cascade coloured,   $N_C$, and lacunar,   $N_L$, components with the equations describing the fractional energy density of nonrelativistic matter and dark energy. 
Such a fitting is exemplified by the arbitrary selection of the following models:
\begin{itemize}
\item The consensus cold dark matter model ($\Lambda$CDM ):  
\begin{align}
     N_L(\tau) &= \frac{ p_m  (1 +  C_z \tau)^3 }{
     p_r (1 + C_z \tau )^4 +
               p_m ( 1 + C_z \tau )^3 +
               p_k ( + C_z \tau )  ^2 + (1 - p_k - p_m - p_r) }
\end{align}
\item The constant $w$ model ($w$CDM) : 
\begin{align}
     N_L(\tau) &= \frac{ p_m  (1 +  C_z \tau)^3 }{
           p_r (1 + C_z \tau )^4 +
               p_m ( 1 + C_z \tau )^3 +
               p_k ( + C_z \tau )  ^2 + (1 - p_k - p_m - p_r) (1 + C_z \tau  )^\omega}
\end{align}
\item The Chevallier–Polarski–Linder  model (CPL): 
\begin{align}
     N_L(\tau) &= 
      \dfrac{ p_m  (1 +  C_z \tau)^3 }{  \splitdfrac{p_r (1 + C_z \tau )^4 +
               p_m ( 1 + C_z \tau )^3 +
               p_k ( + C_z \tau )  ^2 + }{ 
               (1 - p_k - p_m - p_r) (1 + C_z \tau  )^{3(1 + \omega_0 + \omega_a )}
            e{ \frac{-3  \omega_a  C_z \tau }{ 1 + C_z \tau } } }  }
\end{align}
\item The generalised Chaplygin gas model (GCG):
\begin{align}
     N_L(\tau) &= 
      \dfrac{ p_m  (1 +  C_z \tau)^3 }{  \splitdfrac{p_r (1 + C_z \tau )^4 +
               p_b ( 1 + C_z \tau )^3 +
               p_k ( 1 + C_z \tau )  ^2 + }{ 
               (1 - p_k - p_b - p_r) 
               [ ( A_s+ (1 - A_s) (1 + C_z \tau)^{ 3 (1 + \alpha) } ]^{ \frac{1}{1+\alpha} } 
            } }
\end{align}
\item The interacting dark energy model  (IDE):
\begin{align}
	N_L(\tau) &= 
	\dfrac{ p_m  (1 +  C_z \tau)^3 }{  \splitdfrac{p_r (1 + C_z \tau )^4 +
			p_k ( 1 + C_z \tau )  ^2 + }{
			(1 - p_k - p_m - p_r)   ( 1 + C_z \tau )^{3(1+ \omega_x)} +
			\frac{b_m}{\delta + 3 \omega_x}
			[ \delta(1+C_{\tau})^{3(1+\omega_x)} +3 \omega_x(1 + C_z \tau)^{3-\delta }] 
	}}
\end{align}     
\item Early dark energy model (EDE)
\begin{align}
     N_L(\tau) &= 
         \dfrac{ p_m  (1 +  C_z \tau)^3 ( 1 - \Omega_{DE}(z) )}{ 
 p_r (1 + C_z \tau)^4 +
p_m(1 + C_z\tau)^3 +
p_k(1 + C_z\tau)^2 }\\
 \Omega_{DE}(z)  & \equiv  
 ( 
 	(1.0- p_k - p_m  - p_r)- w_e( 1 - (1 + C_z \tau)^{3w_0} ) \nonumber\\
 	&\times [
  (1.0 - p_k - p_m- p_r) +
p_k (1 + C_z \tau)^{-3 w_0 - 1} \nonumber\\
 &+ p_m(1 + C_z \tau)^{-3 w_0}  
+ p_r (1 + C_z \tau)^ {- 3 w_0 + 1} 
]^{-1} \nonumber\\ 
&+
w_e (1 - (1 + C_z \tau)^{3 w_0}).\label{Ode}
\end{align}     
\item Poulin {\it et al.} Early Dark Model (EDEP) 
\begin{align}
N_L(\tau) &=   
 \dfrac{p_m  (1 + C_z \tau )^3}{
 	 \splitdfrac{
  p_r (1 + C_z \tau)^4 +
p_m(1 + C_z \tau)^3 +
p_k(1 + C_z \tau)^2 + }{
(1.0 - p_k - p_m - p_r) +            
\dfrac{ 2 p_a }{\left(\dfrac{ 1 + Z^c}{1 + C_z \tau}\right)^{3(w_n + 1)}  + 1 }  }}
\end{align}     
\end{itemize}
and   $N_C(\tau) = 1 - N_L(\tau)$ for all the cases.

With the exception of   $\Omega_{DE}(\tau)$ - being a function - 
we avoided the conventional use of the sign for the energy densities ($\Omega_x$)
as -- at this point in our discourse --,
the fitting parameters have no physical meaning.
Even so, we have kept the same subscripts to 
maintain a certain parallelism with the original meaning, i.e., 
we deal with parameters 
$p_k$,   $p_m$,   $p_r$ in the   $\Lambda$CDM model. The   $w$CDM adds the parameter   $w$, 
the CPL, GCG, IDE and EDE models include the additional 
parameter sets   $(\omega_0, \omega_a)$,   $(A_s, \alpha)$, 
$(\omega_x, \delta)$ and 	$(\omega_0, \Omega_e)$, respectively, 
while the EDEP includes the parameters   $(p_a,w_n,Z^c)$. 
Used models where inspired by \cite{shi2012} 
where all of them included a contribution from curvature. 
In particular, the EDE model is based on 
\cite{doran2001,doran2006}.  
However the EDEP model was assembled using equations (5) and (15)
in \cite{poulin} and, for fitting purposes, we decided to took the previous triple 
as parameters. 
We incorporated the additional parameter   $C_z$ 
which intents to fit the relationship between the iteration step   $\tau$ and the cosmological redshift   $z$.
Best fit parameters  
using the   \textsc{Turing.jl} package for Bayesian inference with the No-U-Turn sampler \cite{HoffmanGelman} 
are summarised in Table   \ref{table1} and the corner plots 
for the parameters and their covariance are shown in the Figures \ref{Sfig10}-\ref{Sfig15}. 
No corner plot for the GCG model is included as our fitting results were quite unsatisfactory 
(we don't rule out the existence of a better fit parameter set for this case but we haven't found it). 
With the exception of GCG all the models yielded a   $\sigma^2$ measure of the order of   $\sim 10^{-4}$ 
with   $w$CDM reaching a best value of    $\sim 10^{-5}$. 
For all the models   $p_{\Lambda} \sim 0.8$ and   $p_{m}$ is 
in the range   $\sim 0.14 - 0.18$. 
While GCG differs substantially, most models produced   $p_k$   $\,\,\sim 0.00027$, 
excepting EDE with   $p_k \sim 0.005726   $ and EDEP with the only slightly negative value   $p_k \sim  -0.001418$ (positive in the error margin, reported in the Fig. \ref{Sfig15}.
Meanwhile, wCDM, CPL, GCG and IDE reported   $p_r$ negative values in the range    $\sim [-5.8  ,  - 0.2] \times 10^{-3}$, while $\Lambda$CDM, EDE and EDEP models have 
positive   $p_r \sim 0.000137$,   $ 0.000209   $ and   $0.000148$ 
(whose   $\sigma^2$ were   $0.000228$,   $0.000181$ and   $0.000226   $, respectively).
The obtained values for the parameter pairs 
$(\omega_0, \omega_a)$,   $(A_s, \alpha)$ and   $(\omega_x, \delta)$ in the 
CPL, GCG and IDE models are very close to those obtained in \cite{shi2012} 
 - i.e.,   $(-0.966, 0.202)$,   $(0.733,-0.011)$ and   $(-1.001, -0.0043)$, respectively -  
 \footnote{
 \cite{shi2012} makes use of data from the  
 Union2.1 SNe compilation and the WiggleZ BAO measurements, 
together with the WMAP 7-yr distance priors and the observational Hubble data.
}. 
 Given that the only models yielding  
 positive values for   $p_r$ are the   $\Lambda$CDM, EDE and EDEP 
 models we conclude these are the ones best fitting the   $N_L$ data. 
 Remarkably, the best values obtained for the parameter   $C_z$
 for these models are quite close to each other:  
   $0.365496$,   $0.389605$ and   $0.361704$. 
 Thus,  the relationship between 
 iteration steps and the redshift may be 
 written as   $z \sim C_z \tau$.
 Plots with the fitting results for the 
   $\Lambda$CDM,   $w$CDM, CPL, IDE, EDE and EDEP models 
are shown in the Fig. \ref{Sfig16}.

\subsection{$E(z)$ used in the calculations of  the Hubble constant} \label{SM8}

The analysis was restricted to the best behaviour models from  the above, i.e.,   $E(\tau)$, 
\begin{itemize}
\item for the $\Lambda$CDM model  is given by,   
\begin{align}
	E(\tau) &= 		
		p_r (1 + C_z \tau )^4 +
		p_m ( 1 + C_z \tau )^3 +
		p_k (1 + C_z \tau )^2 + (1 - p_k - p_m - p_r) \label{eLCDM},
\end{align}
\item for the EDE model  is given by,   
\begin{align}
	E(\tau) &= 
	\dfrac{		p_r (1 +C_z \tau)^4 +
		p_m(1 + C_z \tau)^3 +
		p_k(1 + C_z \tau)^2 }{1 - \Omega_{DE}(\tau)} \label{eEDE},
\end{align}     
with $\Omega_{DE}(\tau)$	 expressed by eq. (\ref{Ode}), and 
\item for the EDEP model is given by,   
\begin{align}
	E(\tau) &=   
			p_r (1 + C_z \tau)^4 +
			p_m(1 + C_z \tau)^3 +
			p_k(1 + _z \tau)^2 + \nonumber\\
			&\,\,(1.0 - p_k - p_m - p_r) +            
			\dfrac{ 2 p_a }{\left(\dfrac{ 1 + Z^c}{1 + C_z \tau}\right)^{3(w_n + 1)}  + 1} \label{eEDEP}.
\end{align}     
\end{itemize}
The parameter values in these models are those given in Table   \ref{table1}. 

\subsection{$H_0$ error determination} \label{SM9}
 
Native \textsc{Interpolations.jl} doesn't provide built-in confidence intervals. 
To calculate an error estimation for   $H_0$ we took the mean value of the 
propagated error of the equation (\ref{U1}), i.e., 
	\begin{equation} 
	\epsilon(H_0)  \sim  \left< \epsilon (\Upsilon_x) \right > = \dfrac{1}{2} \dfrac{\epsilon (p_{h})}{ p_h } +    
	\dfrac{1}{2}  \left< \dfrac{\epsilon (\mathcal{M}_x)}{\mathcal{M}_x} \right>. \label{eh0}
\end{equation}
In this expression   $\epsilon(\mathcal{M}_x)$ is the 
root mean square deviation, i.e. 
$\epsilon (\mathcal{M}_x) \equiv \sqrt{ \left<  \dfrac{\Delta D_x}{D_x} -  \mathcal{M}^{1/2}_x  \right>^2 }$. 
Note that   $\epsilon \left(\dfrac{\Delta D_x}{D_x}\right) = 0$, as the involved quantities were  measured directly on the fractal. 
 
\newpage

\begin{table}[htb]
	\centering
			\caption{\textbf{
			Best fit parameters obtained using Julia's   \textsc{Turing.jl} Bayesian inference.} Here   $p_{\Lambda} \equiv 1 - p_k - p_m - p_r$; 
			$p_5 = w$ in the   $w$CDM,    $(p_5, p_6)$ are   $(\omega_0, \omega_a)$,   $(A_s, \alpha)$,   $(\omega_x, \delta)$ and 
			$(\omega_0, \Omega_e)$ in the 
			CPL, GCG, IDE and EDE models, respectively; the EDEP model extra parameters are 
			$(p_5, p_6,p_7)=(p_a,w_n,Z^c)$;   $ \sigma^2$ means residuals. In the GCG model   $p_m$ was substituted by 
			the present density parameter of baryonic matter   $p_b = \Omega_b = 0.0451$, according to the WMAP 7yr results 
			\cite{Komatsu2011} following the example in  \cite{shi2012}.	
		}
		\label{table1}

			\begin{tabular}{cccccccc}
			\\
		\hline
				Parameter & $\Lambda$CDM & wCDM & CPL & GCG & IDE & EDE & EDEP\\
		\hline
				$ C_{z}   $ &    $ 0.366038   $  &    $ 0.728667   $  &    $ 0.423091   $  &    $ 0.917701   $  &    $ 0.528297   $ &   $ 0.389605   $&$0.361704$\\
				$ p_k   $ &    $ 0.000268   $  &    $ 0.00027   $  &    $ 0.00027   $  &    $ 0.334792   $  &    $ 0.00027   $ &   $ 0.005726$&$-0.001418$\\
				$ p_m   $ &   $ 0.177603$  &    $0.141675$  &    $0.164878$  &    $ p_b \equiv 0.0451   $  &    $ 0.163759   $ &   $ 0.179941$&$0.180455$\\
				$ p_r   $ &    $ 0.000137   $  &    $ -0.000225   $  &    $ -0.00035   $  &    $ -0.005757   $  &    $ -0.000459   $ &   $ 0.000209$&$0.000148$\\
				$ p_{ \Lambda}   $ &    $ 0.821993   $  &    $ 0.85828   $  &    $ 0.835201   $  &    $ 0.77703   $  &    $ 0.821993   $&$0.814124$&$0.820815$\\
				$ p_5   $ & -- &    $ 0.821767   $  &    $ -1.006021   $  &    $ 0.747408   $  &    $ -0.822978   $  &   $ -0.953692$&$0.219839$ \\
				$ p_6   $ & -- & -- &    $ 0.205858   $  &    $ -0.011034   $  &    $ -0.00453   $ &   $ 0.814125$&$0.565335$\\
				$ p_7   $ & -- & -- &    $  --    $  &    $  --    $  &    $  --    $ &   $  --   $&   $10^6$\\
				$ \sigma^2$ &    $ 0.000228   $  &    $ 0.000097   $  &    $ 0.000126   $  &    $ 0.432337   $  &    $ 0.000134   $ &   $0.000181$&$0.000226$\\
		\hline
			\end{tabular}
\end{table}

\newpage

\begin{table}[htb]
\centering
	\caption{{\bf Best fit parameters for models given by the equation (21)} Bayesian inference with a Markov Chain Monte Carlo 
		and   $300000$ priors.   $H_0$ was obtained from the estimate posterior distributions according to equations 
		(28) and (29 ). 
		Cero errors means they are smaller than   $10^{-2}$.
		Results for   $H_0$ are in units of   $H(\tau)$, i.e. in Km/s/Mpc, as expected. 
	}
	\label{table2}
	
		\begin{tabular}{ccccc}
		\\
		\hline
		Model&   $D_x$ &   $p_h$ &   $p_e$  &   $H_0$\\
		\hline
		$\Lambda$CDM&   $D_{cg}$ &   $116.11 \pm 20.21$ &   $1.54 \pm 0.03$  &   $\boldsymbol{62.79 \pm 5.48}$\\
		$\Lambda$CDM&   $D_{m}$ &   $74.00 \pm 0.00$ &   $1.41 \pm 0.00$  &   $\boldsymbol{70.07 \pm 0.09}$\\
		EDE&   $D_{cg}$ &   $127.87 \pm 2.87$ &   $1.50 \pm 0.04$  &   $78.91 \pm 0.91$\\
		EDE&   $D_{m}$ &   $35.69 \pm 1.23$ &   $1.50 \pm 0.01$  &   $37.57 \pm 0.69$\\
		EDEP&   $D_{cg}$ &   $134.60 \pm 9.99$ &   $1.52 \pm 0.01$  &   $73.55 \pm 2.75$\\
		EDEP&   $D_{m}$ &   $39.81 \pm 1.08$ &   $1.51 \pm 0.00$  &   $36.91 \pm 0.53$\\
		\hline
	\end{tabular}
\end{table}

\begin{figure}[htb]
	\includegraphics[width=.5\textwidth]{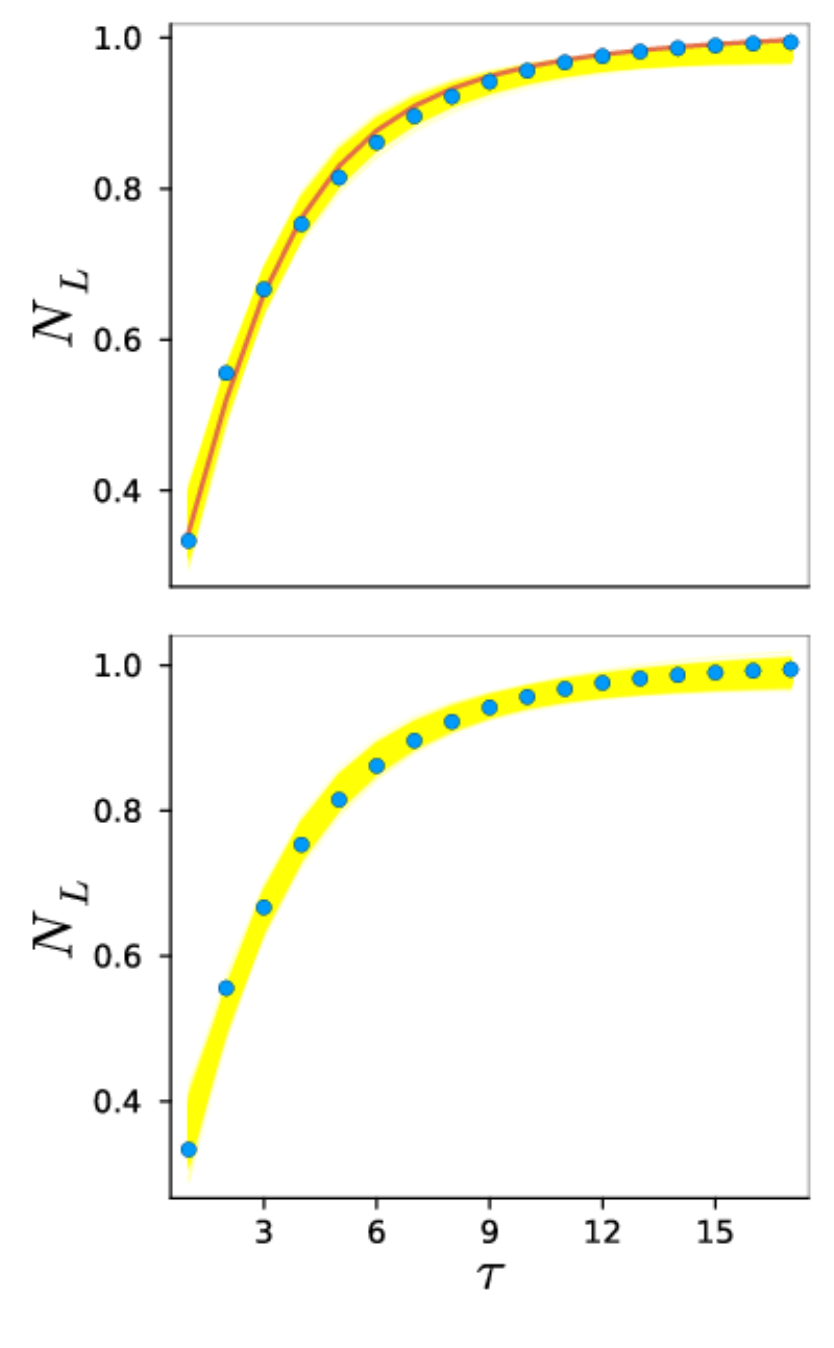}
	\caption{ {\bf  Fitting cosmological models.} (Dots)   $N_L$ vs.   $\tau$ fitted by (top) the 
		$\Lambda$CDM model and (bottom) an Early Dark Energy (EDE) model 
		\cite{doran2001,doran2006,shi2012} using a Markov Chain Monte Carlo analysis. 
		The red line correspond to the fit used in Figure \ref{fig1}B  and  \ref{fig1}D .
		In the    $\Lambda$CDM model 
		$z \sim 0.365496 \,\tau$, while in the EDE model   $z \sim 0.389605 \,\tau$ 
		(   the Appendix ).
	}
	\label{fig2}
\end{figure}

\begin{figure}[htb]
\includegraphics[width=.9\textwidth]{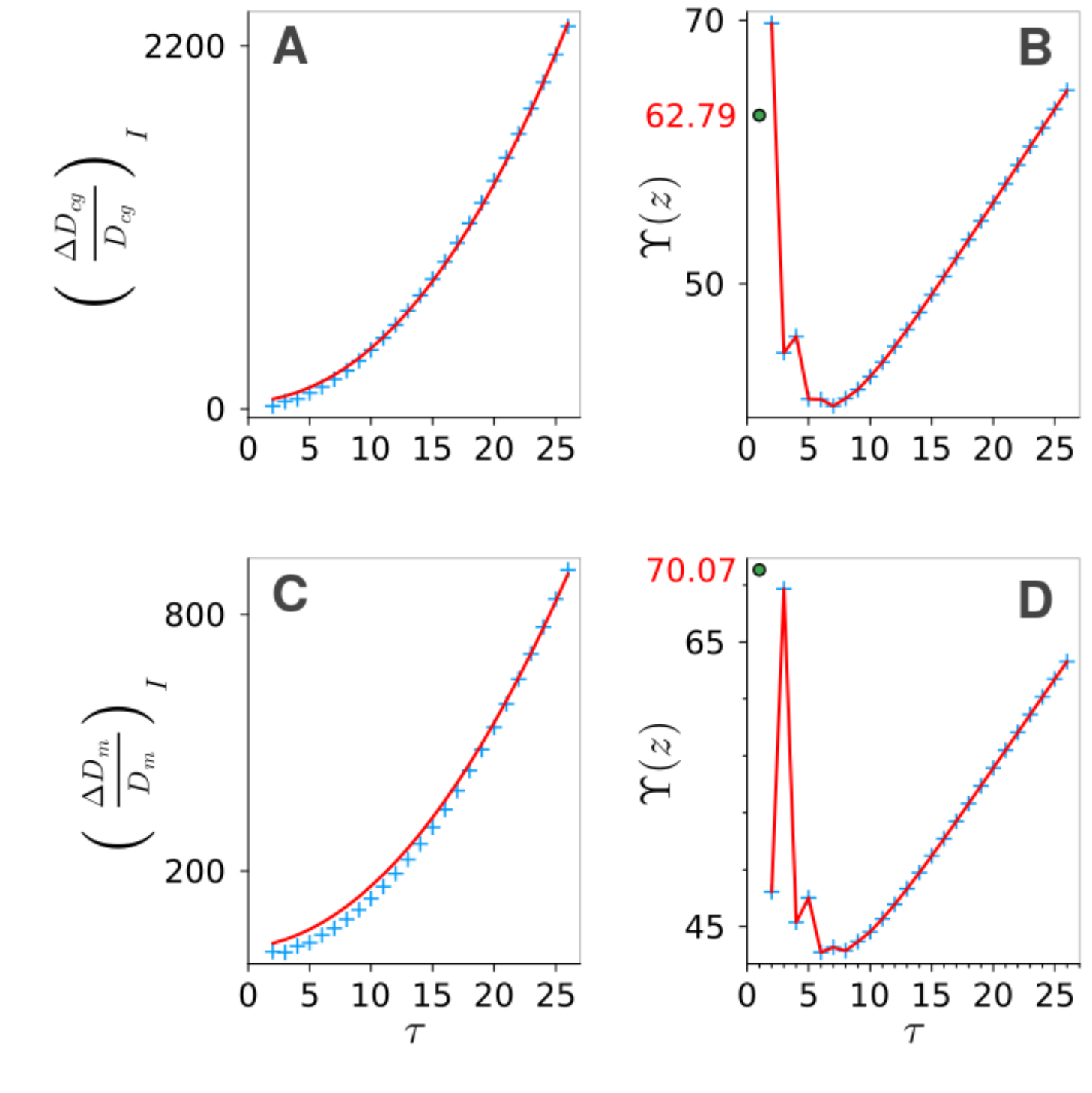}
\caption{{\bf Determination of the Hubble constant from the relative growth of the fractal space.} 
		({\bf A}) and ({\bf C}) show the result of fitting   $\left(\dfrac{\Delta D_{x}}{D_{x}}(\tau) \right)_{I}^2   $ to   $\mathcal{M}(\tau)$,
	using   $E(\tau)$ as given by the   $\Lambda$CDM model (equation (13) in the SI). 
	({\bf B}) and ({\bf D}) show the result of evaluating the function   $\Upsilon(\tau)$ given by (\ref{U1}) (red line) while the plus signs show an interpolating function whose extrapolation to   $\tau \rightarrow 1$ yield the value remarked with the green dot at 	
	({\bf B})   $H_0 = \Upsilon(1) \sim (62.79 \pm 5.48)$ Km/s/Mpc and at  ({\bf D})
	$H_0 = \Upsilon(1) \sim (70.07 \pm 0.09)   $ Km/s/Mpc. 
	Further information about the   $H_0$ error 
	is given in Table  \ref{table2}.
}
\label{fig3}
\end{figure}

\begin{figure}[htb]
	\includegraphics[width=0.9\textwidth]{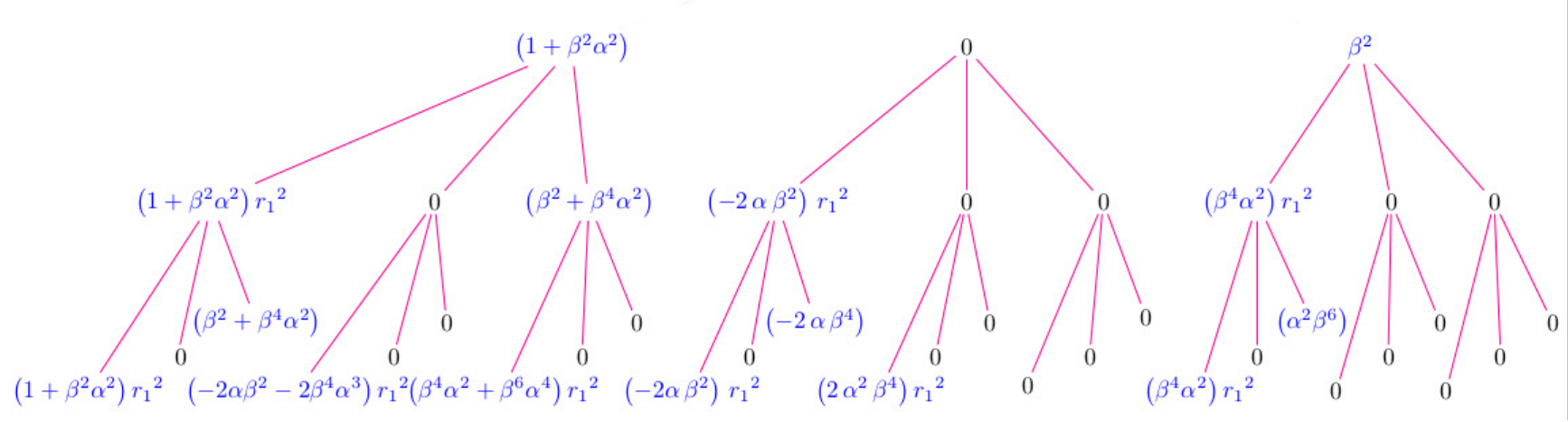}
	\caption{
		{\bf   
		Tree representation for the generation of  the analytical cascade.}
		First steps in the generation of the analytical fractal cascade shown in the main text figure 1.
		This figure fixes a typo in a similar one published before \cite
		{jlc}.
	}\label{Sfig4}
\end{figure}

\begin{figure}[htb]
		\includegraphics[width=0.9\textwidth]{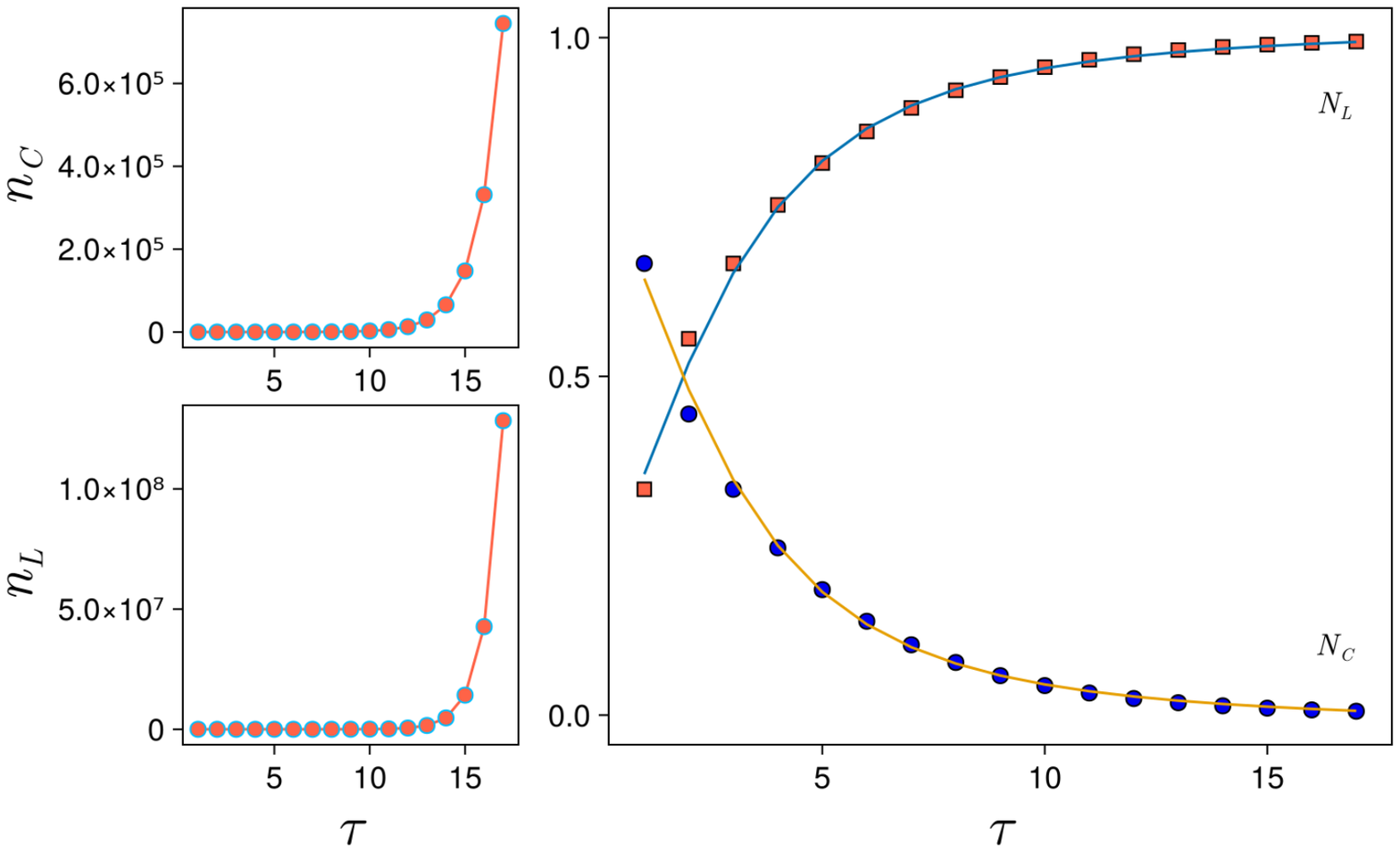}
	\caption{
	         {\bf Behaviour of the cascade components.}
		(Right top:) Number of colored components   $n_C$
		and (Right bottom:) number of lacunar components   $n_L$ in the inverted fractal cascade. 
		(Left:) Fraction of colored   $N_C$ and lacunar   $N_L$ components in the inverted cascade as seen by an observer 
		at the origin. In all the cases the continuous line is the best fit. 
	}\label{Sfig5}
\end{figure}

\begin{figure}[htb]
	\includegraphics[width=\textwidth]{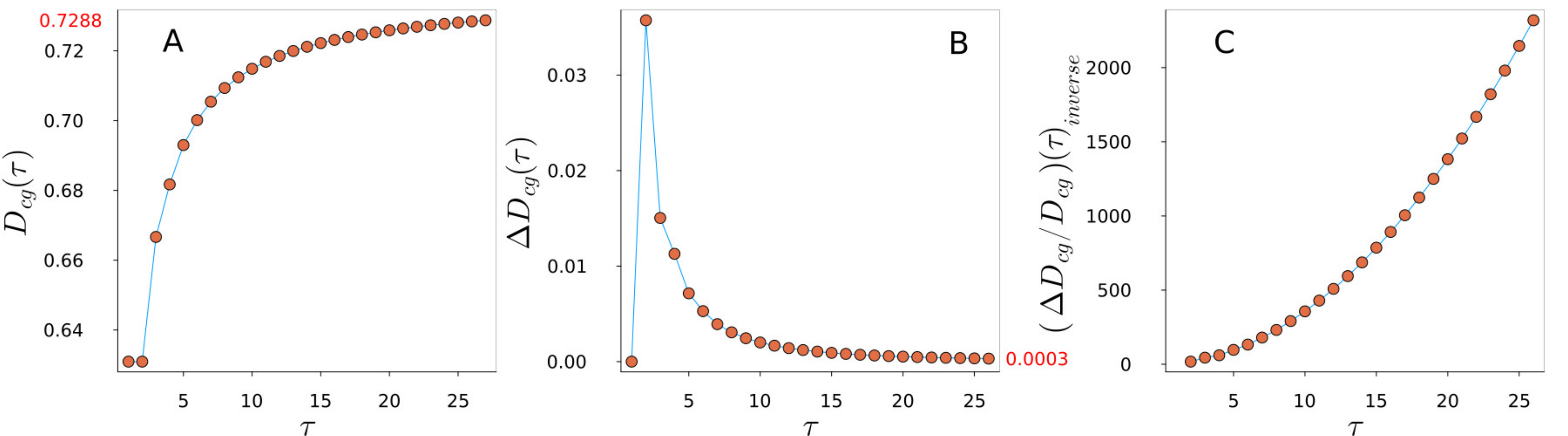}
	\caption{
	{\bf  Course grain dimension.}
		({\bf A})   $D_{cd}(\tau)$ takes the value   $0.7288$ at   $\tau_f=27$, 
		the first two iterates resulted in a constant value as the fractal is beginning to take  shape; 
		({\bf B}) starting from   $\Delta D_{cg}(1)=0$, the consecutive differences,   $\Delta D_{cg}(\tau)$, 
		peaks at   $\tau=2$ and decays to   $\sim 0.0003$ at   $\tau=26$; 
		({\bf C})   
		$\left(\dfrac{\Delta D_{cg}}{D_{cg}} (\tau) \right)_{I} = \dfrac{D_{cg}}{\Delta D_{cg}}(\tau)$, 
		calculated using the represented data in ({\bf A}) and ({\bf B}), 
		note that at   $\tau=1$ the result diverges as   $\Delta D_{cg}(1)=0$, and is not represented (dropped). Data dots are joined by lines. 
	}\label{Sfig6}
\end{figure}   

\begin{figure}[htb] 
	\includegraphics[width=\textwidth]{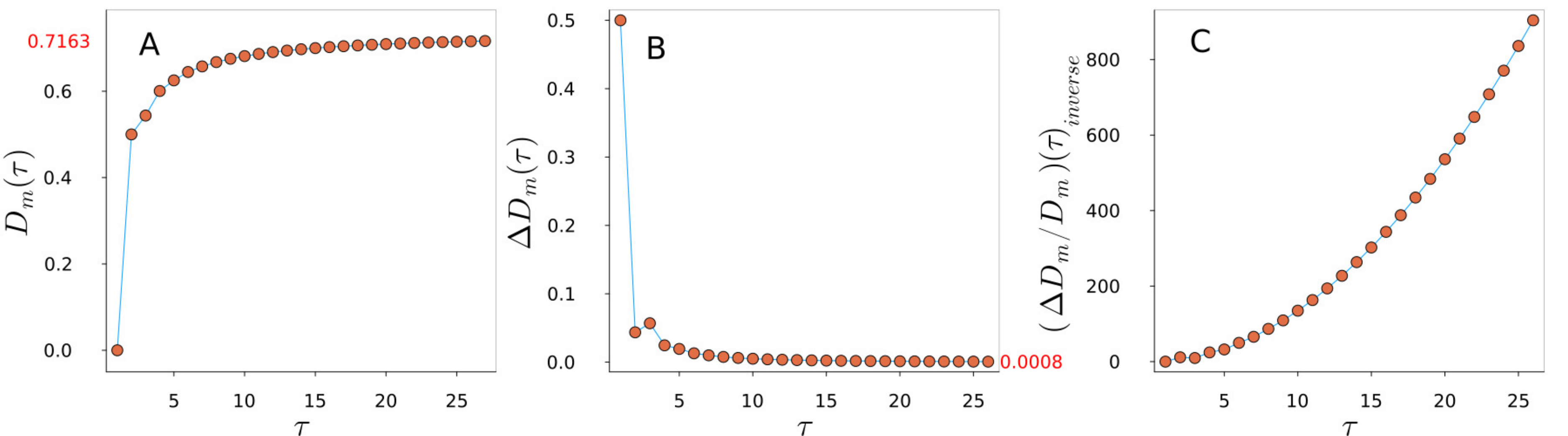}
	\caption{
	{\bf Motifs dimension.}
		({\bf A})   $D_{m}(\tau)$ takes the value   $0.7163$ at   $\tau_f=27$.
		({\bf B}) starting from   $\Delta D_{m}(1) \sim 0.5$, the consecutive differences,   $\Delta D_{m}(\tau)$, 
		decays to   $\sim 0.0008$ at   $\tau=26$. ({\bf C})  
		$\left(\dfrac{\Delta D_{m}}{D_{m}}(\tau)\right)_{I} = \dfrac{D_{m}}{\Delta D_{m}}(\tau)$, 
		calculated using the represented data in ({\bf A}) and ({\bf B}).
		Data dots are joined by lines. 
	}\label{Sfig7}
\end{figure}   

\begin{figure}[htb] 
	\includegraphics[width=\textwidth]{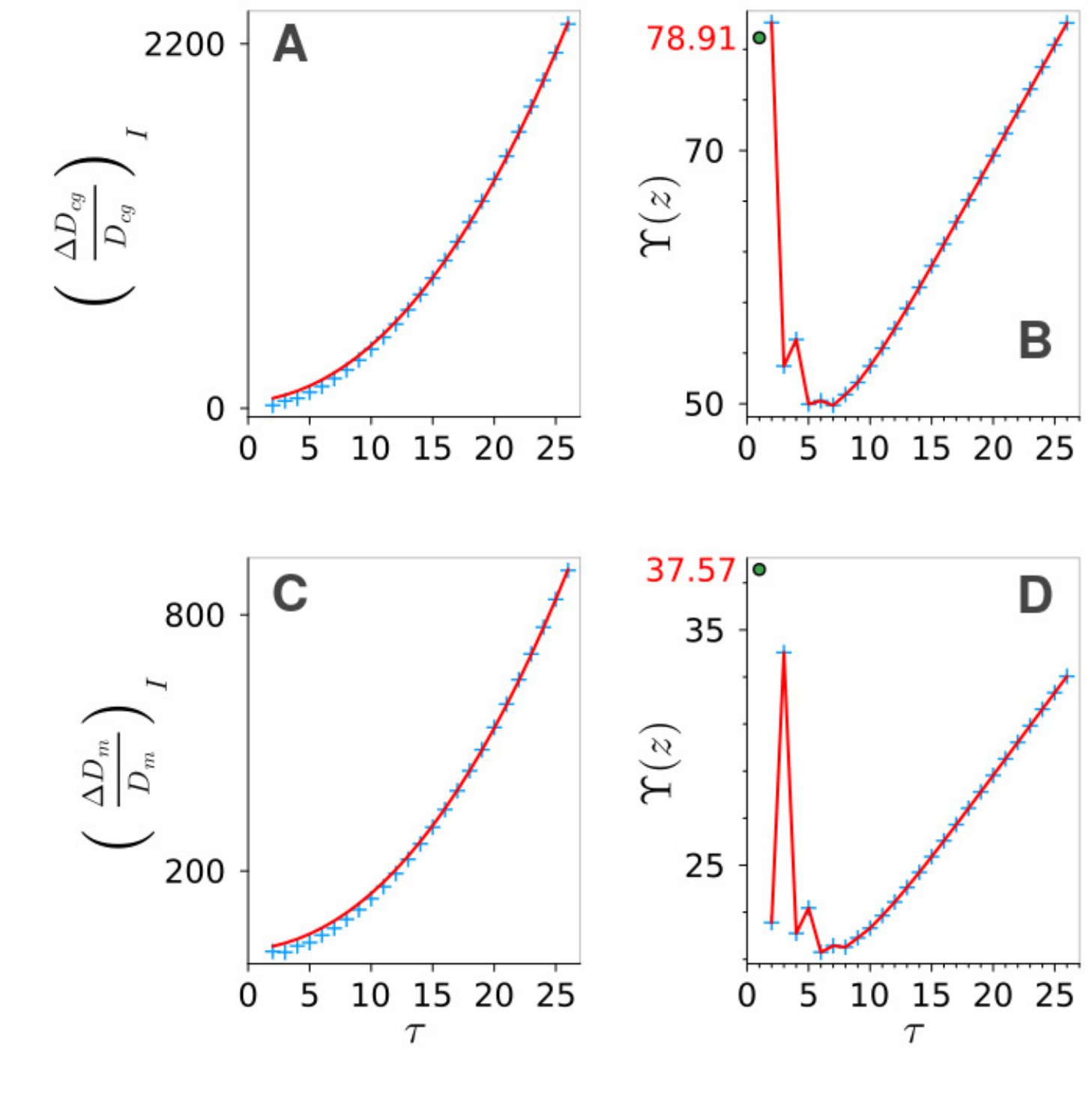}
	\caption{{\bf Determination of the Hubble constant from the relative growth of the fractal space.} 
		Same as in the main text figure 3, 
		but using EDE   $E(t)$, i.e., given by Equation (\ref{eEDE}).
		 In this case 	
		({\bf B})   $H_0 = \Upsilon(1) \sim 78.91 \pm 0.91$ Km/s/Mpc and ({\bf D})
		$H_0 = \Upsilon(1) \sim 37.57 \pm 0.69$ Km/s/Mpc.
	}\label{Sfig8}
\end{figure}

\begin{figure}[htb]  
	\includegraphics[width=\textwidth]{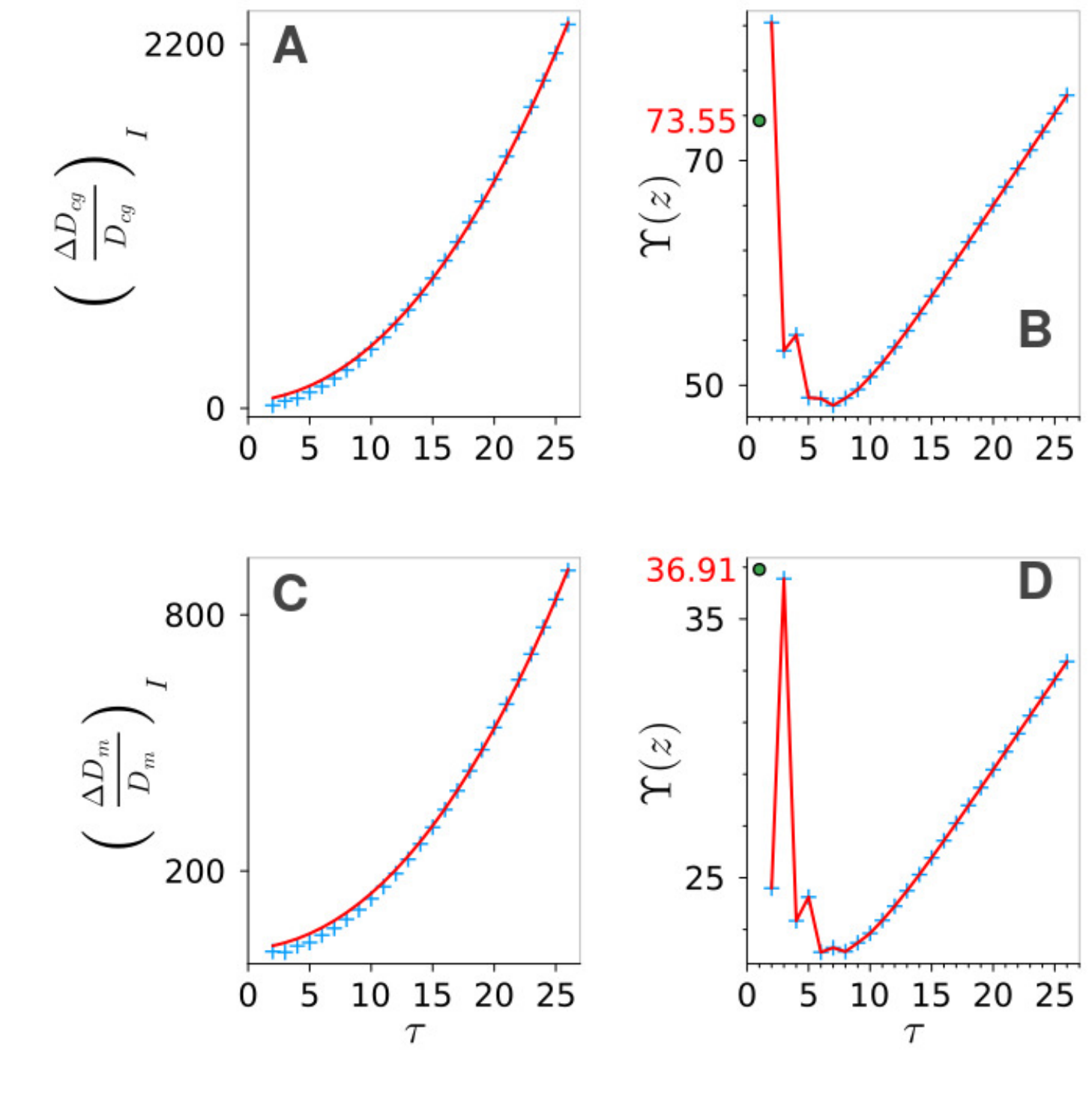}
	\caption{{\bf Determination of the Hubble constant from the relative growth of the fractal space.} 
		Same as above, 
		 but using EDEP   $E(t)$, i.e., given by Equation (\ref{eEDEP}). In this case 	
		({\bf B})   $H_0 = \Upsilon(1) \sim 73.55 \pm 2.75$ Km/s/Mpc and ({\bf D})
		$H_0 = \Upsilon(1) \sim 36.91 \pm 0.53$ Km/s/Mpc.
	}\label{Sfig9}
\end{figure} 

\begin{figure}[htb]  
	\includegraphics[width=\textwidth]{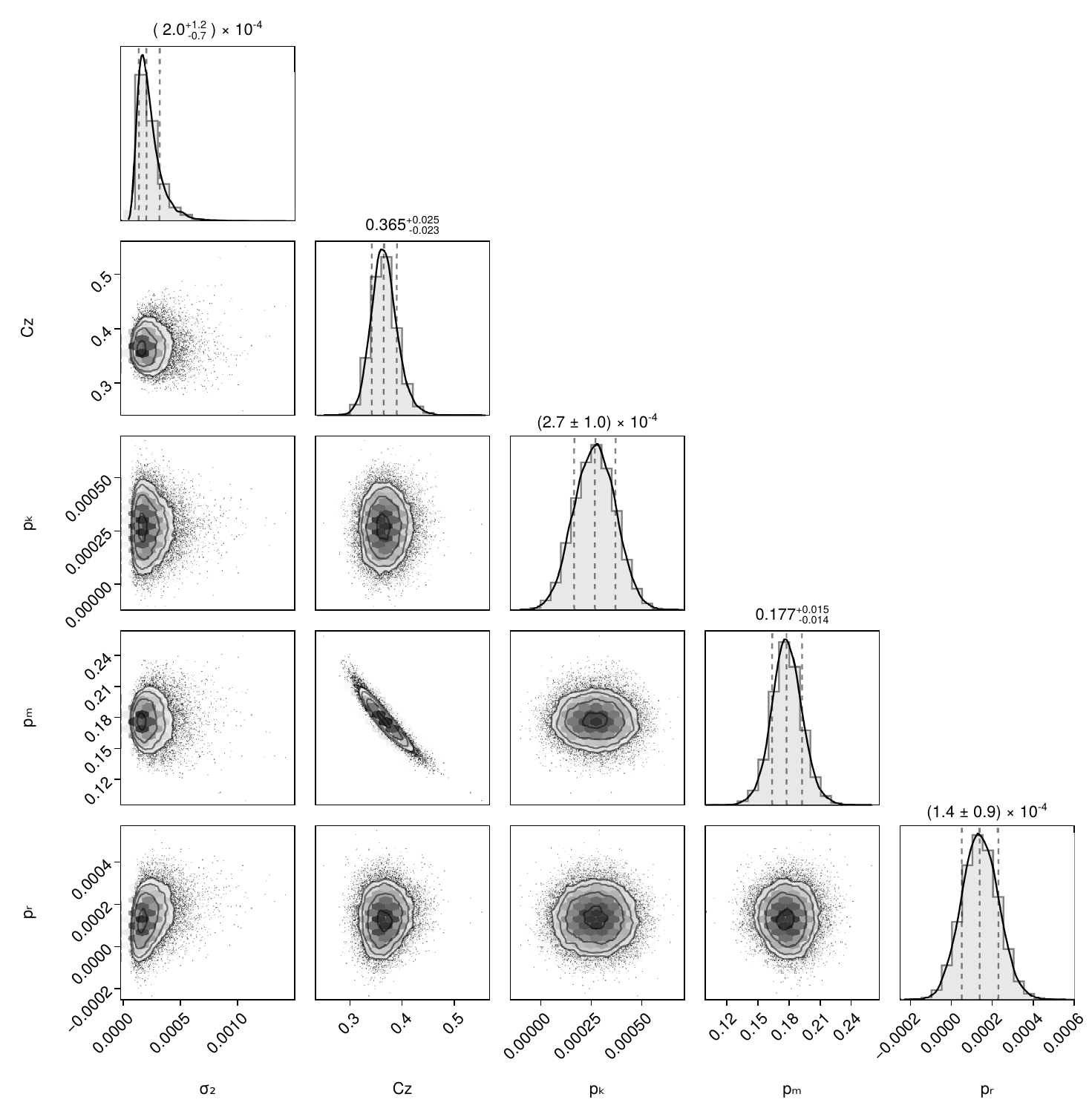}
		\caption{{\bf  Parameters distribution and their covariance for the   $\Lambda$CDM model.}
	}\label{Sfig10}
\end{figure}

\begin{figure}[htb]  
	\includegraphics[width=\textwidth]{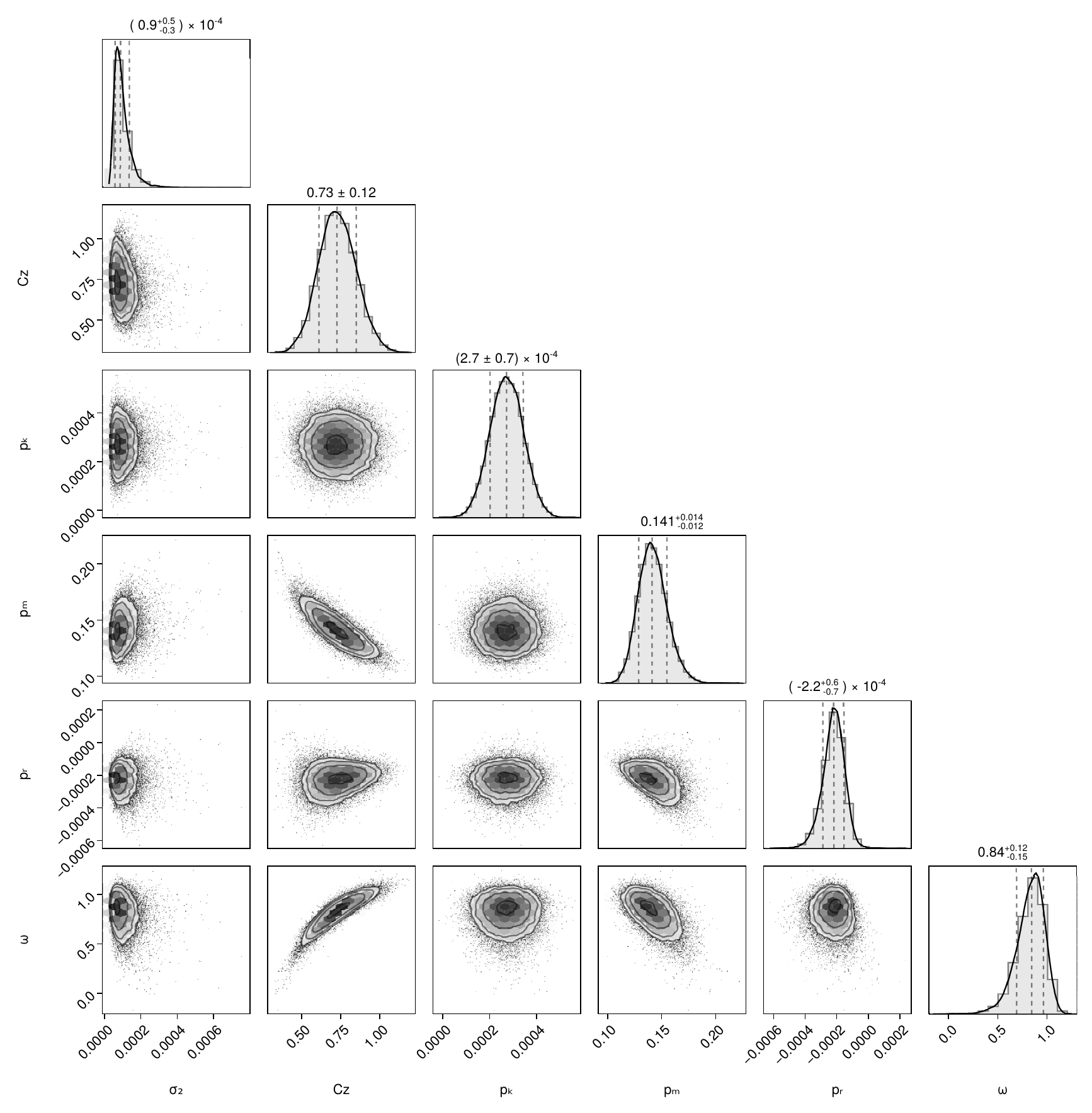}
			\caption{{\bf Parameters distribution and their covariance for the    $w$CDM model.}
	}\label{Sfig11}
\end{figure}

\begin{figure}[htb]  
	\includegraphics[width=\textwidth]{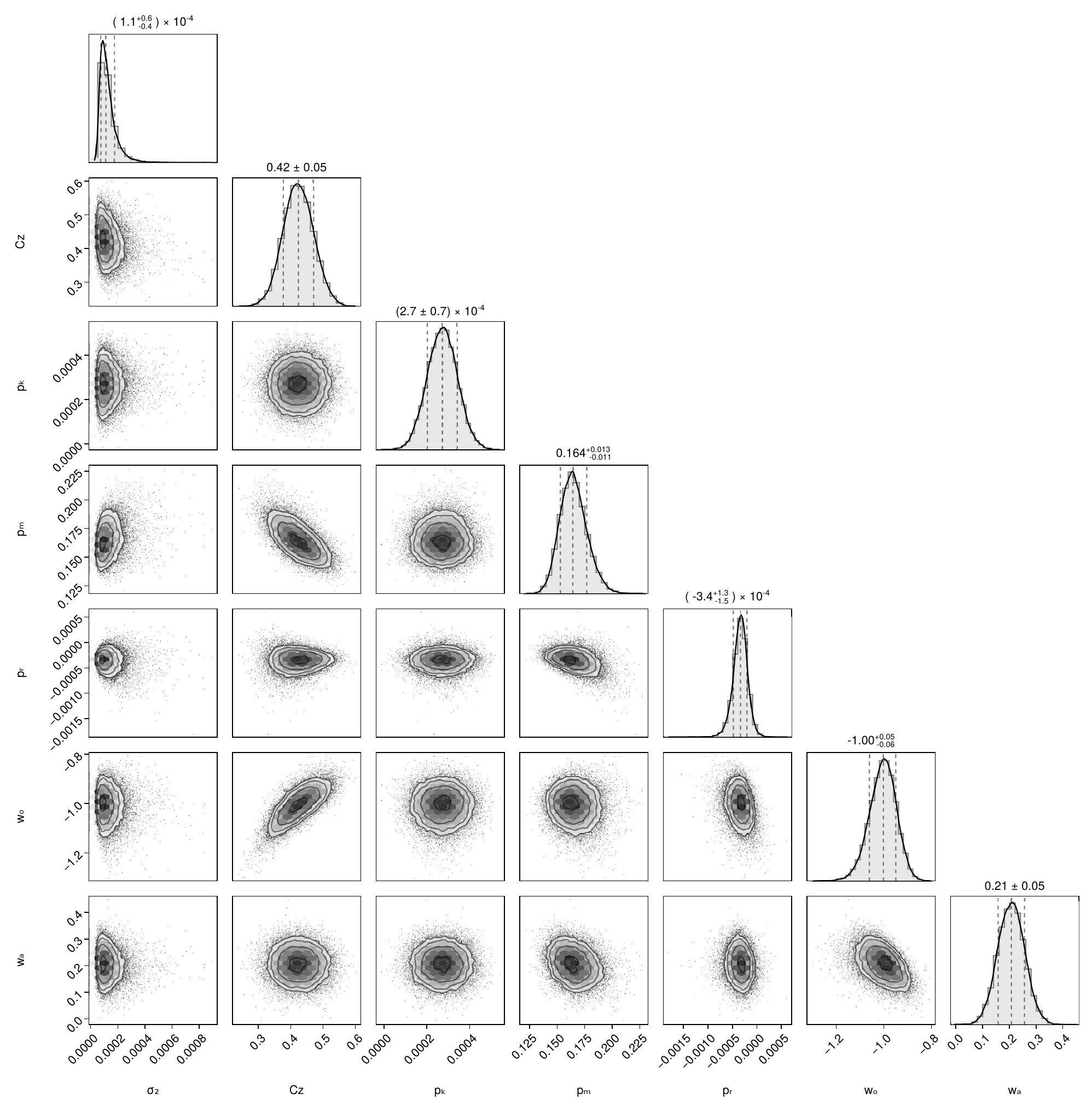}
			\caption{{\bf Parameters distribution and their covariance for the  CPL model.}
	}\label{Sfig12}
\end{figure}

 \begin{figure}[htb]  
 	\includegraphics[width=\textwidth]{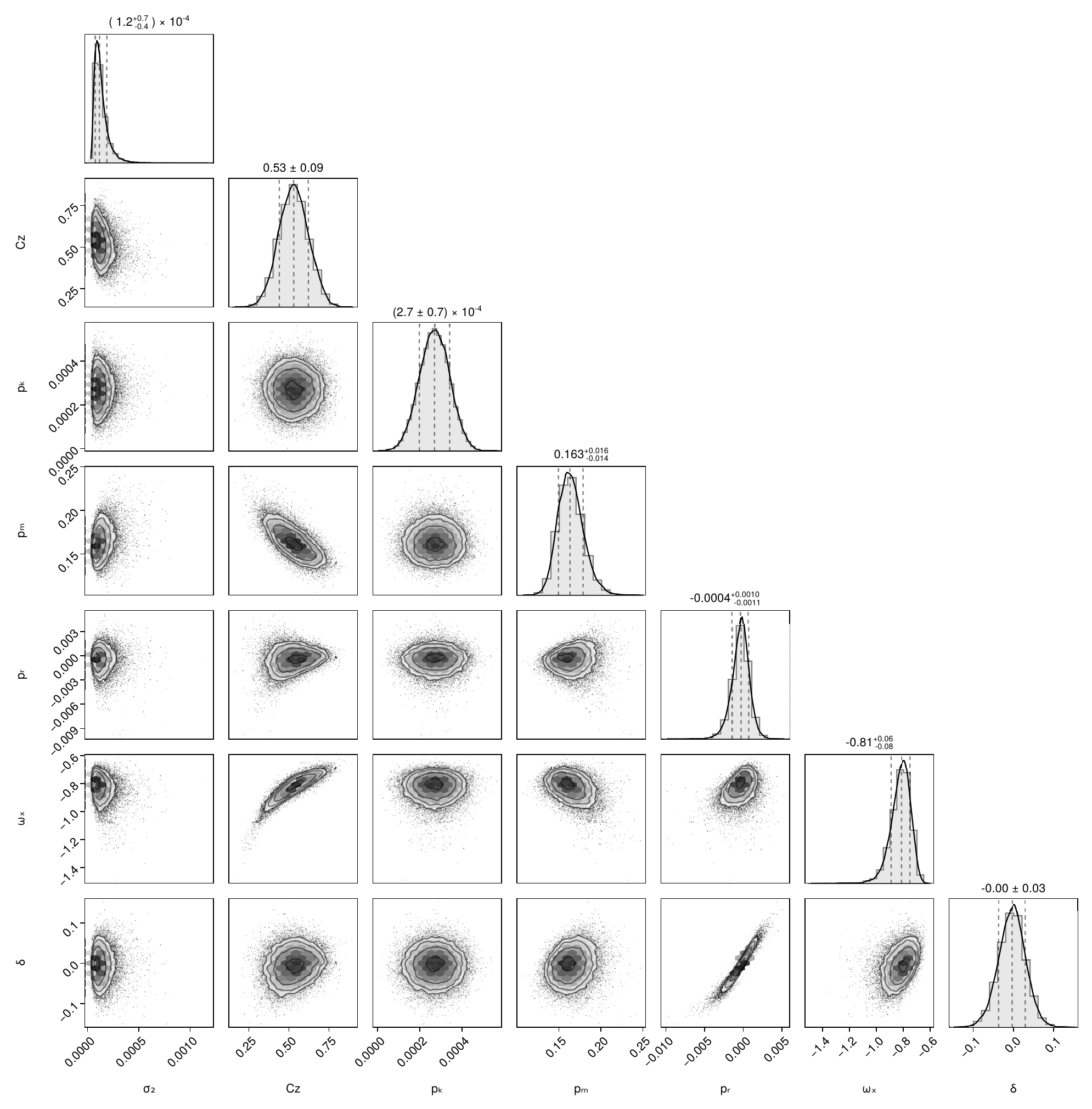}
				\caption{{\bf  Parameters distribution and their covariance for the  IDE model.}
 	}\label{Sfig13}
 \end{figure}
 
 \begin{figure}[htb]  
 	\includegraphics[width=\textwidth]{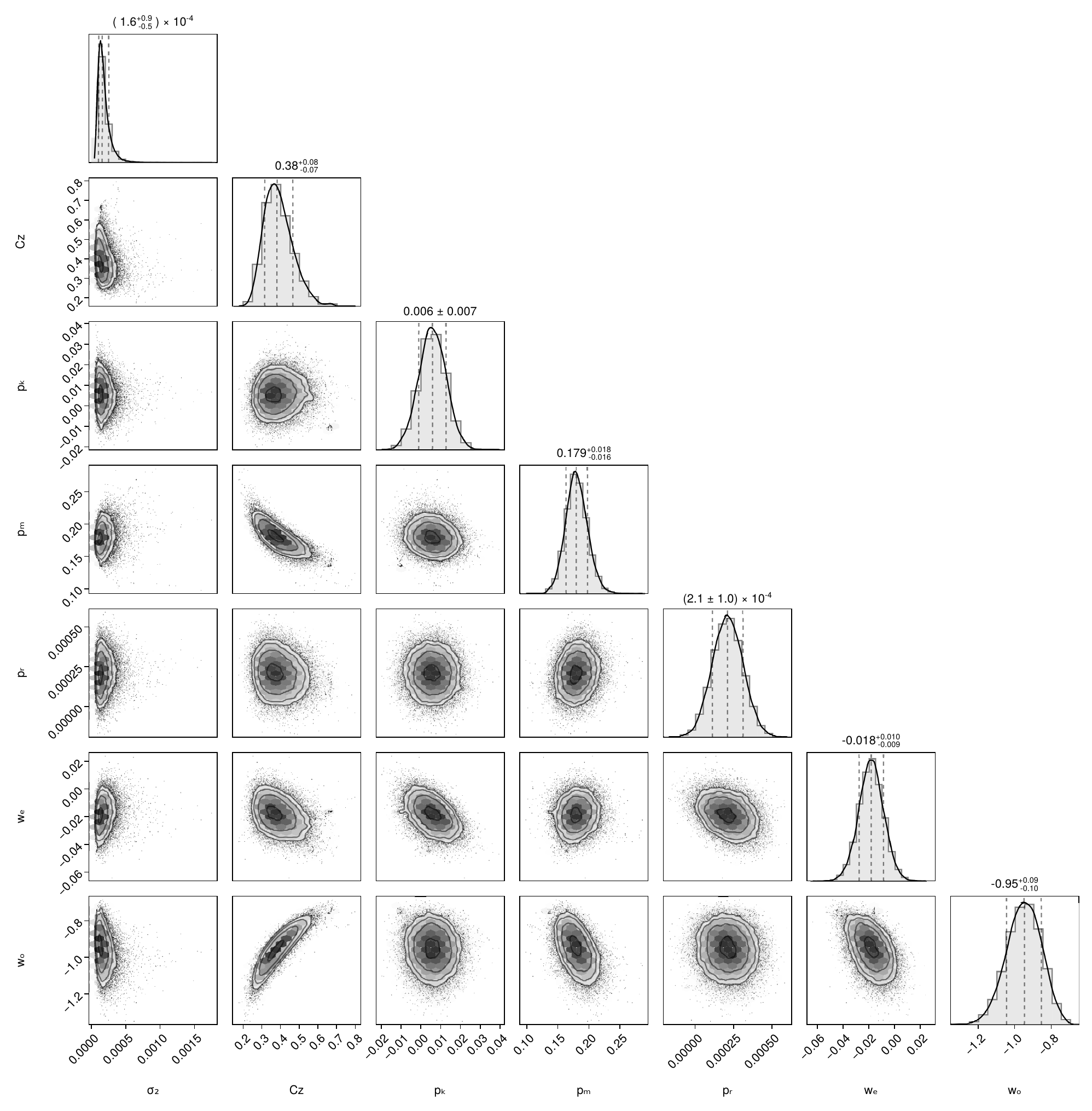}
					\caption{{\bf  Parameters distribution and their covariance for the   EDE model.}
 	}\label{Sfig14}
 \end{figure}

\begin{figure}[htb]  
	\includegraphics[width=\textwidth]{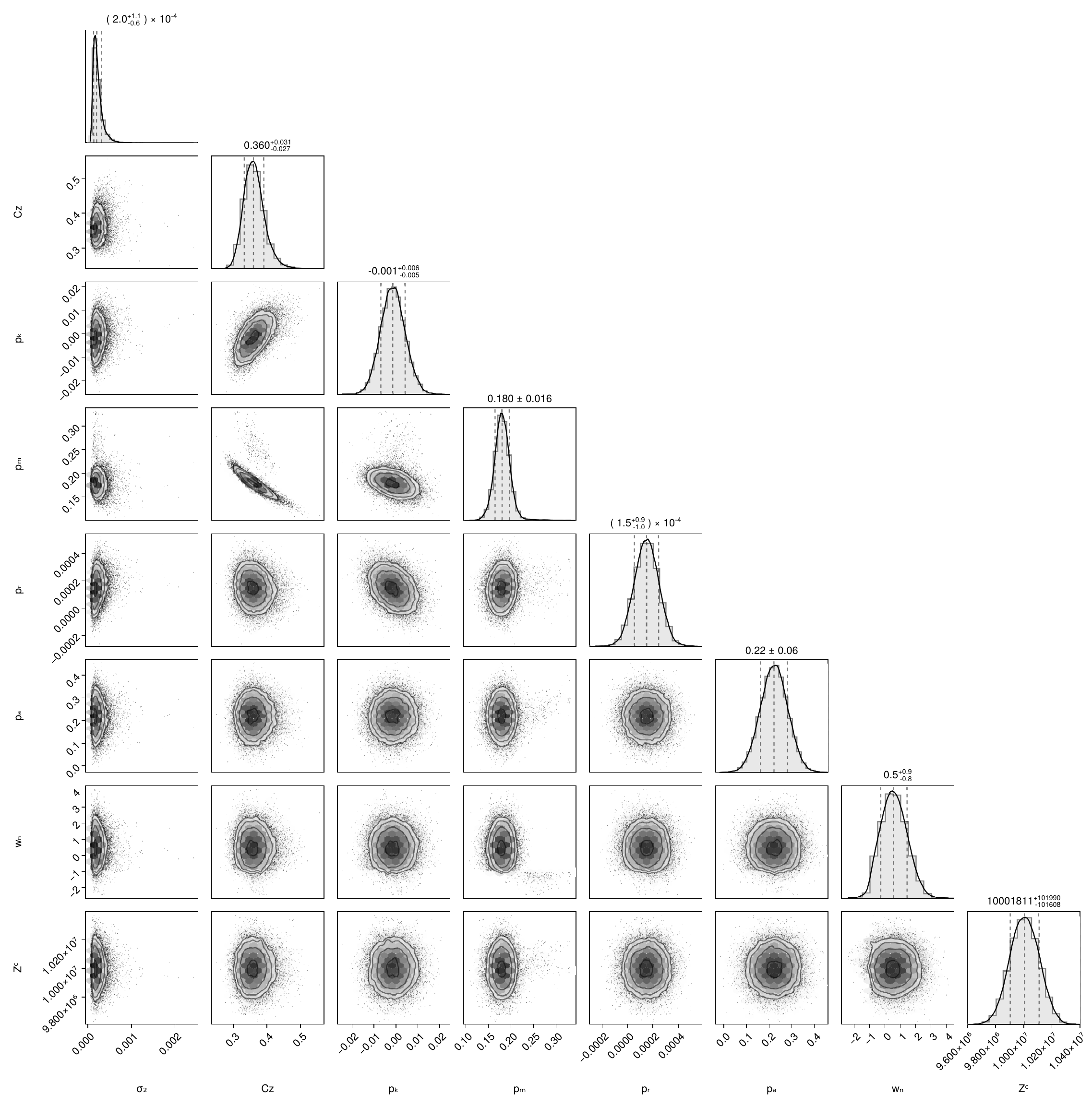}
					\caption{{\bf  Parameters distribution and their covariance for the   EDEP model.}
	}\label{Sfig15}
\end{figure}

\begin{figure}[htb]  
	\includegraphics[width=\textwidth]{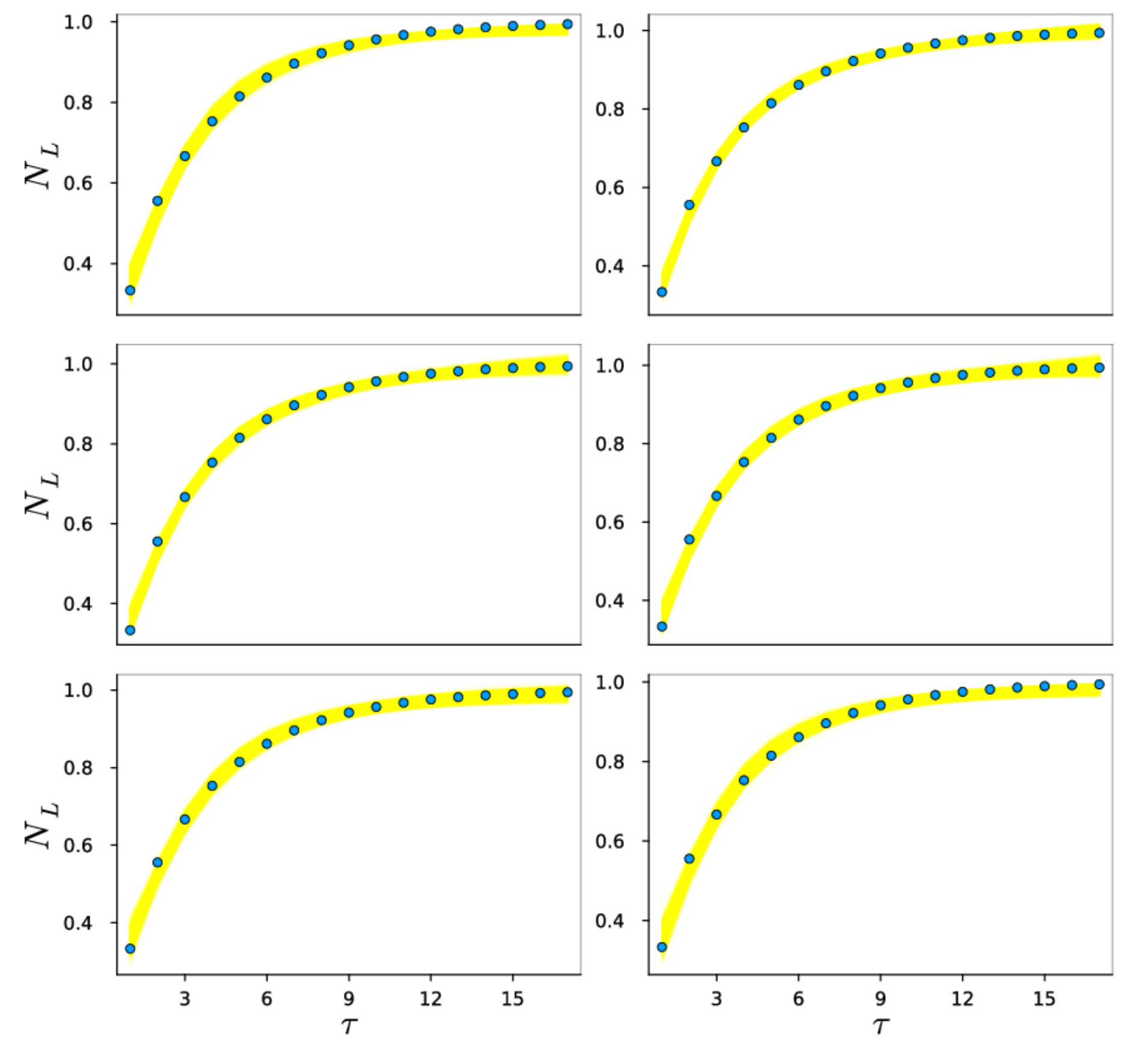}
					\caption{{\bf Fitting   $N_L(\tau)$ with different cosmological models.} From left to right and from top to bottom   $\Lambda$CDM,   $w$CDM, CPL, IDE, EDE and EDEP models fitted using Julia's   \textsc{Turing.jl} Bayesian inference. 
	}\label{Sfig16}
\end{figure}




\bibliography{arxivCosmos.bib}

\end{document}